\documentstyle[11pt,psfig]{article}
\newif
\iffigs
\figstrue     

\def\la{\left\langle}
\def\ra{\right\rangle}
\def\ms1{{\kern .035em s}}

\makeatletter \def\fps@figure{tp} \makeatother
\iffigs
\fi
\def\drawing #1 #2 #3 {
\begin{center}
\setlength{\unitlength}{1mm}
\begin{picture}(#1,#2)(0,0)
\put(0,0){\framebox(#1,#2){#3}}
\end{picture}
\end{center} }

\def\rset{{\rm I\kern -0.2em R}}
\def\un{\hbox{{1\kern -0.25em\raise 0.4ex\hbox{{\scriptsize $|$}}}}}

\def\cset{\hbox{{C\kern -0.55em\raise 0.5ex\hbox{{\tiny $|$}}}}}
\def\nset{\hbox{{I\kern -0.18em N}}}
\def\lsaut{|\![}
\def\rsaut{]\!|}

\begin{document}
\title{Kicked Burgers Turbulence}
\author{J. Bec$^1$, U. Frisch$^1$ \& K. Khanin$^{2-4}$}
\date{}
\maketitle
\vspace*{-0.7cm}
\begin{center}
\begin{tabular}{ll}
$^1$ &Observatoire de la C\^ote d'Azur, Lab. G.D. Cassini,\\
& B.P. 4229, F-06304 Nice Cedex 4, France. E-mail: bec@obs-nice.fr\\
$^2$ &Department of Mathematics, Heriot-Watt University,\\
&Edinburgh EH14 4AS, UK. E-mail: K.Khanin@ma.hw.ac.uk\\
$^3$ &Isaac Newton Institute for Mathematical Sciences,\\
&20 Clarkson Road, Cambridge CB3 0EH, UK.\\
$^4$ &Landau Institute for Theoretical Physics,\\
&Kosygina Str., 2, Moscow 117332, Russia.\\\\
\end{tabular}
\end{center}

\vspace*{-0.3cm}
\centerline{7 April 2000}
\centerline{\it J. Fluid Mech.; in press}
 \begin{abstract}
{\normalsize
\noindent Burgers turbulence subject to a force
$f(x,t)=\sum_jf_j(x)\delta(t-t_j)$, where the $t_j$'s are ``kicking
times'' and the ``impulses'' $f_j(x)$ have arbitrary space dependence,
combines features of the purely decaying and the
continuously forced cases.  With large-scale forcing this ``kicked''
Burgers turbulence presents many of the regimes proposed by E, Khanin,
Mazel and Sinai (1997) for the case of random white-in-time
forcing. It is also amenable to efficient numerical simulations in the
inviscid limit, using a modification of the Fast Legendre Transform
method developed for decaying Burgers turbulence by Noullez and Vergassola
(1994). For the kicked case, concepts such as ``minimizers'' and ``main
shock'', which play  crucial roles in
recent developments for forced Burgers turbulence, become elementary
since everything can be constructed from simple two-dimensional
area-preserving Euler--Lagrange maps.

The main results are for the case of identical deterministic kicks
which are periodic and analytic in space and are applied periodically
in time.  When the space integrals of the initial velocity and of the
impulses vanish, it is proved and illustrated numerically that a space-
and time-periodic solution is achieved exponentially fast. In this
regime, probabilities can be defined by averaging over space and time
periods. The probability densities of large negative velocity
gradients and of (not-too-large) negative velocity increments follow
the power law with $-7/2$ exponent proposed by E {\it et al}.\ (1997) in
the inviscid limit, whose existence is still controversial in the case
of white-in-time forcing. This power law, which is seen very clearly
in the numerical simulations, is the signature of nascent shocks
(preshocks) and holds only when at least one new shock is born between
successive kicks.

It is shown that the third-order structure function over a spatial
separation $\Delta x$ is analytic in $\Delta x$ although the velocity
field is generally only piecewise analytic (i.e.\ between
shocks). Structure functions of order $p\neq 3$ are nonanalytic at
$\Delta x=0$. For even $p$ there is a leading-order term proportional to
$|\Delta x|$ and for odd $p>3$ the leading-order term $\propto \Delta
x$ has a nonanalytic correction  $\propto\Delta x|\Delta x|$ stemming 
from shock mergers. 
}
\end{abstract}
\newpage
\section{Introduction}
\label{s:intro}

The driven Burgers equation
\begin{eqnarray}
&&\partial_t u  +u\partial_x u= \nu \partial_x^2 u +f,\label{burgers}\\
&&u(x,t_0)=u_0(x),
\label{initial}
\end{eqnarray}
with a force $f(x,t)$ has a much richer structure than the decaying
problem with $f=0$. Indeed, by
the Hopf (1950) and Cole (1951) transformation the latter is mapped
into the heat equation, while the former goes into a kind of
imaginary-time Schr\"odinger equation with a potential $F$ such that
$f=-\partial_x F$. Actually, the randomly forced Burgers equation appears in a
number of problems in statistical mechanics. This includes the
Kardar, Parisi \& Zhang (1986) equation for interface dynamics (see also
Barab\'asi \& Stanley 1995) and the problem of directed
polymers in random media (Bouchaud, M\'ezard \& Parisi 1995). 

 From the point of view of turbulence dynamics, the forced Burgers
equation can be used in the same spirit as the forced Navier--Stokes
equation, namely to investigate universality with respect to the
forcing of various statistical properties. For Navier--Stokes
turbulence, when the force is confined to large spatial scales and the
Reynolds number is very high, small-scale (inertial range) statistical
properties are generally conjectured not to depend on the forcing,
except through overall numerical factors. Similar conjectures have
been made for Burgers turbulence with large-scale forcing.  For
example, there is little doubt that, because of the presence of
shocks, all the structure functions of order $p>1$ have a universal
exponent equal to unity (see, e.g., Bouchaud, M\'ezard \& Parisi 1995;
E {\it et al}.\ 1997).  Chekhlov \& Yakhot (1995) were the first to
study the behavior of the probability
density function (pdf) of  velocity increments when the force
is a white-noise process in time and has a $k^{-1}$ spectrum, a case
leading
to anomalous scaling. When
the force is a white-noise process in time but is confined to large
scales, the behavior of the pdf  $p(\xi)$ of the velocity gradient $\xi$ at
large negative values is rather 
controversial. If it is generally believed that this pdf follows a
power law,
\begin{equation}
p(\xi) \propto |\xi|^{\alpha}, \quad {\rm for}\,\, \xi \to -\infty,
\label{pdfalpha}
\end{equation}
the conjectured values of $\alpha$ differ markedly.  Polyakov (1995)
and Boldyrev (1997), using a field-theoretical operator product
expansion, predict $\alpha=-5/2$; E {\it et al}.\ (1997), using a
semi-heuristic approach in which nascent shocks (preshocks) are key,
predict $\alpha=-7/2$; Gotoh \& Kraichnan (1998), using a
Fokker--Planck equation approach, predict $\alpha=-3$; more recent
work by Kraichnan (1999) favors $\alpha=-7/2$. E \& Vanden Eijnden
(1999, 2000) develop a probabilistic formalism that copes with the
delicate problems arising in the limit of vanishing viscosity when
shocks are present, prove that $\alpha < -3$ and make a good case for
$\alpha=-7/2$.  The question of the correct law for the case of
white-noise forcing remains however open (we shall come back to this
in \S\ref{s:conclusion}). There are simpler situations for which the
arguments in favor of $\alpha=-7/2$, originally developed by E {\it
et al}.\ (1997), can be made rigorous, such as one- and
multi-dimensional decaying Burgers turbulence with smooth random
initial conditions (Bec \& Frisch 2000; Bec, Frisch \& Villone 2000).

It may be thought that numerical experimentation on the
one-dimensional forced Burgers equation should be able to easily
obtain the correct scaling laws. This is actually rather difficult if
one tries to use standard numerical schemes of the kind also
applicable to the Navier--Stokes equation, such as spectral methods
with dissipation explicitly taken into account. Indeed, it has been
shown by Gotoh \& Kraichnan (1998; see also Gotoh 1999) that, in the
presence of a small but finite viscosity, there is a range of large
negative values of the gradient for which the pdf $p(\xi)\propto \nu
|\xi|^{-1}$, which decreases rather slowly at large $|\xi|$. This is a
direct consequence of the hyperbolic-tangent internal structure of
shocks. As a result, the behavior of the pdf at negative values of
$\xi$ smaller in absolute value is contaminated and may display power
laws somewhat shallower than predicted by the theory in the inviscid
limit. See, for example, figure~3 of Gotoh \& Kraichnan (1998) in
which a power-law range with exponent -3 is seen over a little more
than one decade of $\xi$-values. Such artefacts will disappear if much
higher resolution is used (Gotoh and Kraichnan were using between
$2^{17}$ and $2^{20}$ collocation points). It is, however, much more
efficient to use alternative numerical schemes where one works
directly with the inviscid limit. We shall see that such schemes,
which were previously used only for the decaying case can be readily
extended to the forced case and are particularly well suited for the
case of kicked Burgers turbulence, in which the force is concentrated
at discrete times.

The paper is organized as follows. Section~\ref{s:kicked} introduces
the general concept of kicked Burgers turbulence and gives an explicit
representation of the solution which is a simple generalization of the
so-called ``minimum representation'' for the decaying
case. Section~\ref{s:minimizer} presents the concepts of
``minimizers'' and ``main shocks'' for the case when the dynamics are
started at $t_0=-\infty$.  Section~\ref{s:numerical} presents the Fast
Legendre Transform numerical scheme for space-periodic kicked Burgers
turbulence with spatially smooth forcing. The next two sections are
about the case of deterministically kicked Burgers turbulence when the
kicks are periodic in space and time.  Section~\ref{s:periodic}
presents numerical results on exponential convergence to a periodic
solution and shows how this is related to properties of minimizers
(the rigorous results on convergence to a unique periodic solution are
derived in the Appendix). Section~\ref{s:statistical} presents the
main results about pdf's of velocity derivatives and increments
(\S\ref{s:pdf}) and about structure functions
(\S\ref{s:structure}). Section~\ref{s:conclusion} presents concluding
remarks and some possible extensions to Navier--Stokes turbulence.

\section{The inviscid limit for kicked Burgers turbulence}
\label{s:kicked}

We shall be concerned here with the initial-value problem for the
one-dimensional Burgers equation (\ref{burgers}) when the force is
concentrated at discrete times:
\begin{equation}
f(x,t) = \sum_j f_j(x)\,\delta(t-t_j),
\label{kickforce}
\end{equation}
where $\delta$ is the Dirac distribution and where both the
``impulses'' $f_j(x)$ and the ``kicking times'' $t_j$ are prescribed
(deterministic or random). The kicking times are ordered and form a
finite or infinite sequence. In this paper the impulses are always
taken smooth and  acting only at large scales. Newman \& McKane (1997)
have used similar kicking, but confined to small scales, in a context where
the forced Burgers equation is used for the study of directed
polymers. Kraichnan (1999) has
considered a simple model in which there are non-smooth impulses
creating directly sawtooth profiles in the velocity.  The precise
meaning we ascribe to the Burgers equation with this  forcing is that
at time $t_j$, the solution $u(x,t)$ changes discontinuously by the
amount $f_j(x)$
\begin{equation}
u(x,t_{j+})= u(x,t_{j-})+f_j(x),
\label{discon}
\end{equation}
while, between $t_{j+}$ and $t_{(j+1)-}$ the solution evolves according to
the unforced Burgers equation. Without loss of generality, we can
assume that the earliest  kicking time is $t_{j_0}=t_0$, provided we 
set $f_{j_0}=u_0$ and $u(x,t)=0$ for $t<t_0$.

It is clear that any force $f(x,t)$ which is continuously acting in
time can be approximated in such a way by choosing the kicking times
sufficiently close.

We shall also make use of the formulation in terms 
of the velocity potential $\psi(x,t)$ and the force potentials $F_j(x)$
\begin{equation}
u(x,t) = - \partial_x \psi(x,t),\qquad f_j(x)= - {d\over dx} F_j(x).
\label{psiF}
\end{equation}
 The velocity potential satisfies
\begin{eqnarray}
&&\partial_t \psi = \frac{1}{2} (\partial_x \psi)^2 +
\nu \partial_{xx} \psi + \sum_j F_j(x)\,\delta(t-t_j),
\label{BEP}\\
&&\psi(x,t_0)=\psi_0(x),
\label{initpsi}
\end{eqnarray}
where $\psi_0(x)$ is the initial potential.

As it is well-known, the solution to the {\em unforced\/} Burgers
equation with positive viscosity $\nu$, has an explicit integral
representation obtained by Hopf (1950) and Cole (1951) which can be
used to investigate the limit of vanishing viscosity.  We are here
exclusively interested in this limit. Generically, shocks appear then
after a finite time from smooth initial data.  (The correct solution
may also be obtained by solving the inviscid equation with a
variational formulation and the condition that, at a shock, the right
velocity is less than the left velocity (Lax 1957; Oleinik 1957).) Use
of Laplace's method then leads to the following ``minimum
representation'' for the potential in the limit of vanishing viscosity
(henceforth always understood) which relates the solutions at any two
times $t>t'$ between which no force is applied:
\begin{equation}
\psi (x,t) = -\min_y \left[\frac{(x-y)^2 } {2(t-t')}- \psi(y,t')\right].
\label{MAXunforced}
\end{equation}
It is known that, when $t'$ is the initial time, the position $y$
which minimizes (\ref{MAXunforced}) is the Lagrangian coordinate
associated to the Eulerian coordinate $x$. The map $y\mapsto x$ is
called the Lagrangian map. By expanding the quadratic term it is
easily shown that the calculation of $\psi(\cdot,t)$ from
$\psi(\cdot,t')$ is equivalent to a Legendre transformation. For
details on all these matters, see She, Aurell \& Frisch (1992) and
Vergassola {\it et al}.\ (1994).

We now turn to the {\em forced\/} case with impulses applied at the
kicking times $t_j$. Let $t_{J(t)}$ be the last such time before
$t$. (Henceforth we shall often just write $t_J$.)  Using
(\ref{MAXunforced}) iteratively between kicks and changing the
potential $\psi(y,t_{j+1})$ discontinuously by the amount $F_{j+1}(y)$
at times $t_{j+1}$, we obtain
\begin{eqnarray} 
\psi(x,t) &=& -\min_{y_J,y_{J-1},\ldots,y_{j_0}}
\left[A(j_0;x,t;\{y_j\})
- \psi_0(y_{j_0})\right],
\label{MAXkicked}\\
A(j_0;x,t;\{y_j\})&\equiv& {(x-y_J)^2\over 2(t-t_J)}
+\sum_{j=j_0}^{J-1}\left[ {(y_{j+1}-y_j)^2\over 2(t_{j+1}-t_j)} 
-F_{j+1}(y_{j+1}) \right],
\label{defA}
\end{eqnarray}
where $A(j_0;x,t;\{y_j\})$ is called the action.

For the Burgers equation with  a 
continuous-in-time force deriving from a potential $F(x,t)$, E
{\it et al}.\ (1997, 2000) give a  minimizer representation of the solution:
\begin{eqnarray} 
\psi(x,t) &=& -\min_{y(\cdot)}
\left[A(t_0;x,t;y(\cdot))
- \psi_0(y(t_0))\right],
\label{MAXcontin}\\[1ex]
A(t_0;x,t;y(\cdot))&\equiv&\int_{t_0}^t \left[{\dot y^2(s)\over 2}-
F(y(s),s)\right]\,ds,
\label{defAcont}
\end{eqnarray}
where the minimum is taken over all curves $y(s)$ satisfying
$y(t)=x$. Note that this  representation, which actually goes back to work by
Oleinik (1957) on general conservation laws, is just a continuous
limit of (\ref{MAXkicked})-(\ref{defA}), obtained by taking
$t_j=j\,\Delta t$ and letting $\Delta t\to 0$.

Returning to the case of kicked Burgers turbulence, from (\ref{MAXkicked}), we
shall now introduce the concept of ``minimizers'' and ``main shock''
(\S\ref{s:minimizer}).  Eq.~(\ref{MAXkicked}) will also be our
starting point for the numerical method (\S\ref{s:numerical}). For the
rest of the paper, we shall assume that {\em the force potential and
the initial condition are periodic in the space variable.} For
convenience, the period is  taken to be unity in the theory, while
$2\pi$-periodicity is assumed in numerical studies.

\subsection{Minimizers and main shocks}
\label{s:minimizer}

For the case of the kicked Burgers equation with an initial condition
at $t_{j_0}$ a ``minimizing sequence'' associated to $(x,t)$ is defined as a
sequence of $y_j$'s ($j=j_0,j_0+1,\ldots, J(t)$) at which the r.h.s. of
(\ref{MAXkicked}) achieves its minimum.  Differentiating the action
(\ref{defA}) with respect to the $y_j$'s one gets  necessary
conditions for such a  sequence, which can be written as a sequence of 
(Euler--Lagrange) maps
\begin{eqnarray}
&&v_{j+1} = v_j + f_j(y_j),\label{map1}\\
&&y_{j+1} = y_j + v_{j+1}(t_{j+1}-t_j)= y_j + (v_j +
f_j(y_j))(t_{j+1}-t_j), 
\label{map2} 
\end{eqnarray}
where 
\begin{equation}
v_j \equiv {y_j-y_{j-1}\over t_j-t_{j-1}}.
\label{defvj}
\end{equation}
These equations must be supplemented by the initial and final conditions: 
\begin{eqnarray}
v_{j_0} &=& u_0(y_{j_0}),\label{mapinit}\\
x &=& y_J + v_{J+1}(t-t_J).
\label{mapfin}
\end{eqnarray}
It is easily seen that $u(x,t)= v_{J+1} = (x-y_J)/(t-t_J)$. Observe that 
the ``particle velocity'' $v_j$  is the velocity of the fluid
particle which arrives at $y_j$ at time $t_j$ and which, of course, has
remained unchanged since the last kick (in Lagrangian
coordinates). Eq.~(\ref{map1}) just expresses that the  particle
velocity changes by $f_j(y_j)$ at the the kicking time $t_j$.

Note that (\ref{map1})-(\ref{map2}) define an area-preserving and
(explicitly) invertible map.

The presence of a force, deterministic or random, allows a formulation
of the Burgers equation in the semi-infinite time interval $]-\infty,
t]$ without fully specifying the initial condition $u_0(x)$ but only
its (spatial) mean value $\la u\ra \equiv \int _0 ^1
u_0(x)dx$.  Heuristically, this
follows from the observation that, for a force of zero spatial mean value,
as assumed here, $\la u\ra$ is a first integral, and
hence does not depend on time, while all the other information
contained in the initial condition is eventually forgotten.

Actually, the construction of the solution in a semi-infinite time
interval is done by extending the concept of minimizing sequence to
the case of dynamics starting at $t_0=-\infty$. For a semi-infinite
sequence $\{y_j\}$ ($j\le J$), let  us define the action
$A(-\infty;x,t;\{y_j\})$ by (\ref{defA}) with $j_0=-\infty$. Such a
semi-infinite sequence will be  called a
``minimizer'' (or  ``one-sided minimizer'') if it minimizes this action
with respect to any  modification of a finite number of
$y_j$'s. Specifically, for any other sequence
$\{{\hat y_j}\}$ which coincides with $\{y_j\}$ except for finitely
many $j$'s (i.e.\ ${\hat y_j}=y_j$, $j\le J-k, k\ge 0$), we require
\begin{equation}
A(J-k;x,t;\{\hat y_j\})\ge A(J-k;x,t;\{y_j\}).
\label{inegalaction}
\end{equation}

Of course, the Euler--Lagrange relations (\ref{map1})-(\ref{map2})
still apply to such minimizers. Hence, if for a given $x$ and $t$ we
know $u(x,t)$ we can recursively construct the minimizer 
$\{y_j\}$ backwards in time by using the inverse of
(\ref{map1})-(\ref{map2}) for all $j<J$ and the final condition -- now
an initial condition -- (\ref{mapfin}) with $v_{J+1} = u(x, t)$. This is
well defined except where $u(x,t)$ has a shock and thus more than one
value.  Actually, solutions in a semi-infinite time interval are
constructed from minimizers and not the other way round.

One way to construct minimizers is to take a sequence of initial
conditions at different times $t_0\to -\infty$. At each such time some
initial condition $u_0(x)$ is given with the only constraint that it
have the same prescribed value for $\la u\ra$. Then, (finite)
minimizing sequences extending from $t_0$ to $t$ are constructed for
these different initial conditions. This sequence of minimizing
sequences has limiting points (sequences themselves) which are
precisely minimizers (E {\it et al}.\ 2000). The uniqueness of such
minimizers, which would then imply the uniqueness of a solution to
Burgers equation in the time interval $]-\infty,t]$, can only be shown
by using additional assumptions, for example for the case of
time-periodic forcing (\S\ref{s:periodic} and Appendix).

If $\la u\ra=0$, the sequence $\{y_j\}$ minimizes the action
$A(-\infty;x,t;\{y_j\})$ in a stronger sense. Consider any sequence
$\{{\hat y_j}\}$ such that, for some integer $P$ we have ${\hat y_j} =
y_j+P$, $j\le J-k, k\ge0$ and which differs arbitrarily from $\{y_j\}$
for $j>J-k$. (In other words, in a sufficiently remote past the hatted
sequence is just shifted by some integer multiple of the spatial
period.)  We then have
\begin{equation}
A(-\infty;x,t;\{\hat y_j\})\ge A(-\infty;x,t;\{y_j\}).
\label{inegalperiodic}
\end{equation}
Indeed, for $\la u\ra=0$, the velocity
potential for any initial condition is itself periodic.  In this case
a particle can be considered as moving on the circle $S^1$ and its
trajectory is a curve on the space-time cylinder. The $y_j$'s are now
defined modulo~1 and can be coded on a representative $0\le y_j<
1$. The Euler--Lagrange map (\ref{map1})-(\ref{map2}) is still valid
provided (\ref{map2}) is defined modulo~1. 

The condition of minimality implies now that $y_j$ and $y_{j+1}$ are
connected by the shortest possible straight segment. It follows that
$\vert v_{j+1}\vert = \rho(y_j, y_{j+1})/(t_{j+1}-t_j)$, where
$\rho$ is the distance on the circle between the points $y_j,
y_{j+1}$, namely $\rho(a,b)\equiv \min \{|a-b|, 1 -|a-b|\}$. 
Hence, the action $A$ can be rewritten in terms of
cyclic variables:
\begin{equation}
A(-\infty;x,t;\{y_j\}) = {\rho ^2(x,y_J)\over 2(t-t_J)}+
\sum _{j<J}\left [{{\rho ^2(y_{j+1}, y_j)}\over {2(t_{j+1}-t_j)}} -
F_{j+1}(y_{j+1})\right ].
\label{actionz}
\end{equation}

We now introduce the concept of ``global minimizers'' (or ``two-sided
minimizers'') limiting ourselves to the case $\la u\ra=0$ for
simplicity. We first observe that any minimizer $\{y_j, j\le J\}$
can be continued for all $j\ge J$ and hence times $t'>t$ by using the
system (\ref{map1})-(\ref{map2}). However, this procedure,
when extended too far in time, will not usually generate a minimizer
associated to time $t'$. Nevertheless, for any time $t$ there always
exist positions $x$ such that the corresponding minimizers $\{y_j,
j\le J(t)\}$ can be continued to the bilateral sequence $\{y_j,
-\infty < j <+\infty \}$ while keeping the minimizing property. Such
global minimizers 
correspond to trajectories of fluid particles which, from $t=-\infty$
to $t=+\infty$, have never been absorbed in a shock.

We then observe that any shock existing at time $t$ can be continued for
all times $s>t$\,: shocks can merge but they cannot otherwise
disappear. However, since new shocks can be produced, it is not always
possible to trace back an existing shock for arbitrary times $s<t$.  A
shock with this property of having always existed in the past is
called a ``main shock''. 

In E {\it et al}.\ (2000) it is shown that for the case of random
forcing which is $1$-periodic in space and white noise in time rather
than impulsive, the solution of Burgers equation in $]-\infty, t]$ is
unique for $\la u\ra=0$. It is also shown that at time $t$, the set
of points $x\in S^1$ with more than one minimizer, that is shock
locations, is finite and that the main shock and the global minimizer
are unique. The global minimizer forms a hyperbolic trajectory of the
Euler-Lagrange equations and all other minimizers approach the global
one, as $t\to -\infty$ exponentially fast. In particular the two
minimizers associated to the main shock approach the global minimizer
in the remote past but it may be shown that they do so from opposite
directions on the circle $S^1$.  We shall see below in
\S\ref{s:periodic} that the same picture holds in the case of generic
time-periodic kicking.

\subsection{A Fast Legendre Transform numerical method}
\label{s:numerical}

The numerical method used here solves the kicked Burgers equation
directly in the inviscid limit. The basic ideas are very simple: at
each kicking time $t_j$ the potential is changed by the amount
$F_j(x)$; between two successive kicks (or between the last kick and
the output time) the decaying Burgers equation is solved using the
minimum representation (\ref{MAXunforced}); this procedure is repeated
as many times as the number of kicks  between the initial time
and the output time.

Specifically, the space periodic interval, here taken to be
$[0,2\pi[$, is discretized on a regular grid of $N$ collocation points
$x_k\equiv 2k\pi/N$. For each of these positions, (\ref{MAXunforced})
is used to determine the potential at a time $t$ in terms of the
potential just before the last kicking time $t_J$, time at which the
potential has discontinuously changed by $F_J(x)$. Hence, we have~:
\begin{equation}
\psi_k(t) \equiv \psi (x_k,t) = -\min_\ell \left [ \frac{(x_k-y_\ell)^2 }
{2(t-t_J)} -\left(\psi_\ell(t_{J_-}) + F_J(y_\ell)\right ) \right ],
\label{MAXflt}
\end{equation}
where $\psi_\ell(t_{J-})$ is the value of the velocity potential at
$x=y_\ell$, just before the kick.  We note $y_{\ell_k}$ the minimizing
position corresponding to $x_k$. This procedure is, in principle, 
applied recursively, starting from $t_{j_0}$.

The problem is that naive application of (\ref{MAXflt}) yields an
algorithm with $O(N ^2)$ operations between two successive kicks.
She, Aurell \& Frisch (1992) observed that the minimizing position $y$
is actually a monotonic non-decreasing function of $x$. This is indeed
a simple consequence of the convexity of the parabolic term involved
in (\ref{MAXunforced}). Hence, the determination of the $y_{\ell_k}$'s can
be performed using a binary-subdivision search which requires only
$O(N\log_2 N)$ operations.  This kind of algorithm is known under the
name Fast Legendre Transform (FLT), since the minimum representation
is equivalent to a Legendre transform (Hopf 1950). (An even faster
algorithm requiring only $O(N)$ operations has been developed by
Trussov (1996).)

We use an adaptation of the method of Noullez and Vergassola (1994;
see also Vergassola {\it et al}.\ 1994) who developed an FLT algorithm using
a binary-subdivision search combined with a reorganization of the
search, permitting the use of very low in-core storage.  We first
determine the minimizing $y_{\ell_0}$ for the point $x_0\equiv0$. As
the velocity-potential at the time $t_{J_+}$ is periodic, it is easy
to show that $y_{\ell_0}$ is within the interval $[-\pi,+\pi[$. The
minimizing location corresponding to $x_N\equiv 2\pi$ is then given by
periodicity and reads $y_{\ell_N}=y_{\ell_0}+2\pi=y_{\ell_0+N}$.  The
search for all the other minimizing locations $y_{\ell_k}$ can then be
restricted to indices $\ell_k$ such that $\ell_0 \le \ell_k \le
\ell_0+N$. We then compute $y_{\ell_{N/2}}$, corresponding to
$x=\pi$. We can then further subdivide the $x$-interval by considering
$k= N/4$ and $k=3N/4$, for which the corresponding $\ell_k$'s satisfy
$\ell_0 \le \ell_{N/4} \le \ell_{N/2} \le \ell_{3N/4} \le
\ell_0+N$. At the next stage, we compute $y_{\ell_{N/8}}$,
$y_{\ell_{3N/8}}$, $y_{\ell_{5N/8}}$ and $y_{\ell_{7N/8}}$ for which
we need $\ell_0$, $\ell_{N/4}$, $\ell_{N/2}$ and $\ell_{3N/4}$ as
search boundaries.  We repeat this subdivision procedure $\log_2 N$
times to obtain the $N$ values of $y_{\ell_k}$.

The method just described is optimal for non-smooth solutions of the
kind considered by Vergassola {\it et al}.\ (1994) who had initial
conditions of Brownian type. For the case of smooth solutions
considered here, a more accurate determination of the solution is
required to obtain reliable results on space derivatives of the
velocity. We now describe an improvement of the method allowing to
calculate first- and second-order derivatives. We observe that in
(\ref{MAXflt}), when the discrete location $y_\ell$ is replaced by an
arbitrary real number $y$, the minimum, for a given $x_k$ is, in
general, not achieved exactly on the grid at $y_{\ell_k}$, but at a
neighboring location $y(k)$ within less than one mesh. This location
satisfies $x_k=y(k)+(t-t_J)\left[u\left(y(k),t_{J-}\right)+f_J(y(k))
\right]$, obtained by requiring that the derivative of the r.h.s. of
(\ref{MAXflt}) vanish. For this, the velocity $u(y,t_{J-})$ and the
force $f_J(y(k))$  are Taylor expanded to second order to obtain the
improved location $y(k)$. When required, the first and second space
derivatives of the velocity $u(x,t)$ are calculated, not from the
potential by finite differences, but by using exact expressions of
these derivatives in terms of the Lagrangian map from the preceding
kicking time, which holds for the unforced Burgers equation (E {\it et
al}.\ 1997; Bec \& Frisch 1999).

This FLT, implemented on a grid of $2^{17} \approx 10^5$ collocation
points takes about 1\,s of CPU on a 100\,MFlops computer. Without
use of the binary-subdivision search the CPU time would be several
thousands times larger. Among the other
advantages of FLT is that no viscosity is needed and that the solution
can be calculated directly at the required output times without need
to obtain it at many intermediate times (other than the kicking
times).

\section{Deterministic periodic kicking}
\label{s:periodic}

 From now on we shall consider exclusively the case where the kicking
is periodic in both space and time. Specifically, we assume that 
the force in the Burgers equation is given by
\begin{eqnarray}
f(x,t) &=& g(x)\sum_{j=-\infty}^{+\infty} \delta(t-jT),\label{forceperiodic}\\
 g(x)&\equiv& -{d\over dx}G(x),
\label{defg}
\end{eqnarray}
where $G(x)$, the kicking potential,  is a deterministic function of $x$ which is periodic and
sufficiently smooth (e.g.\ analytic) and where $T$ is the kicking
period. The initial potential $\psi_{\rm init}(x)$ is also assumed
smooth and periodic. This implies that the initial velocity integrates
to zero over the period. (The case where this assumption is relaxed
will be considered briefly in the Conclusion in relation with the
Aubry--Mather theory.)

The numerical experiments reported hereafter have been made with the 
kicking potential
\begin{equation}
G(x) = {1\over3}\sin 3x +\cos x,
\label{defG}
\end{equation}
and a kicking period $T=1$. Other experiments where done with (i)
$G(x)= -\cos x$ and (ii) $G(x)= (1/2) \cos (2x) -\cos x$. The former
potential produces a single shock and no preshock. As a consequence it
displays no $-7/2$ law in the pdf of gradients. (In
this paper, we limit ourselves to cases presenting at least one
new shock between successive kicks.)  The latter potential
gives essentially the same results as reported hereafter but has an
additional symmetry which we avoided by the choice (\ref{defG}).

 The number of collocation points chosen for our simulations is
generally $N_x=2^{17}\approx 1.31\times 10^5$, with a few simulations
done at $N_x=2^{20}$ (for the study of the relaxation to the periodic
regime presented below). Since our numerical method allows us to go
directly to the desired output time (from the nearest kicking time)
there is no need to specify a numerical time step. However, in order
to perform temporal averages, e.g.\ when calculating pdf's or
structure functions, without missing the most relevant events (which
can be sharply localized in time) we need sufficiently frequent
temporal sampling. We have taken for the total number of output times
$N_t\approx 1000$ chosen such that the increment between successive
output times is roughly the two-thirds power of the mesh (this is
related to the structure of preshocks, see \S\ref{s:pdf}).

\begin{figure}
\iffigs 
\centerline{\psfig{file=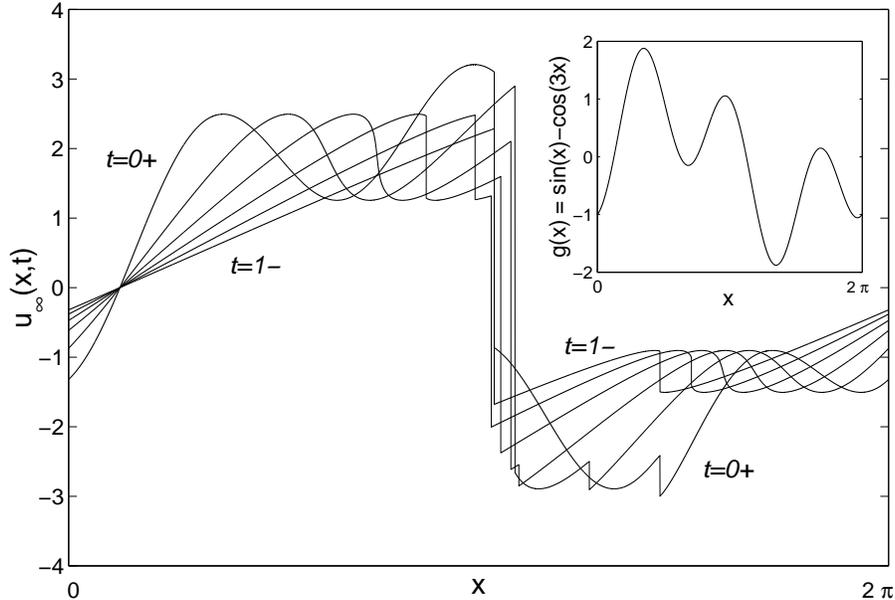,width=12cm}}
\else\drawing 65 10 {snapshots of the periodic solution}
\fi
\vspace{2mm}
\caption{Snapshots of the velocity for the unique time-periodic
solution corresponding to the kicking force $g(x)$ shown in the upper
inset; the various graphs correspond to six output times equally
spaced during one period. The origin of time is taken at a
kick. Notice that during  each period, two new shocks are born  and two
mergers occur.}
\label{f:global-evol}
\end{figure}

Figure~\ref{f:global-evol} shows snapshots of the time-periodic
solution at various instants. 
It is seen that shocks are always present (at least two) and that at
each period two new shocks are born at $t_{\star 1}\approx 0.39$ and
$t_{\star 2}\approx 0.67$. There is one main shock which
remains near $x=\pi$  and which collides with the newborn shocks
at $t_{c1}\approx 0.44$ and  $t_{c2}\approx 0.86$.
Figure~\ref{f:shock-position} shows the  evolution of the  positions of
shocks  during one period.

\begin{figure}
\iffigs 
\centerline{\psfig{file=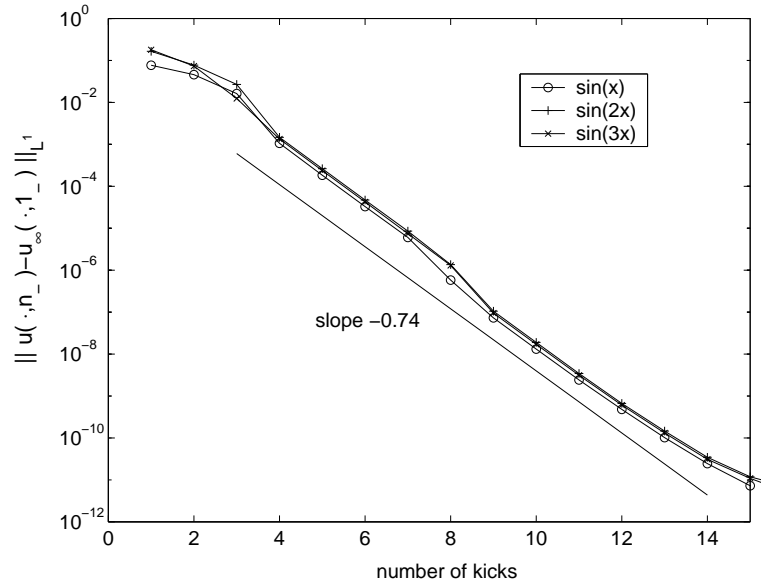,width=10cm}}
\else\drawing 65 10 {exponential relaxation}
\fi
\vspace{2mm}
\caption{Exponential relaxation to a time-periodic solution 
for three different initial velocity data as labelled. The horizontal axis 
gives  the  time elapsed since $t=0$.}
\label{f:relax-expo}
\end{figure}

\begin{figure}
\iffigs 
\centerline{\psfig{file=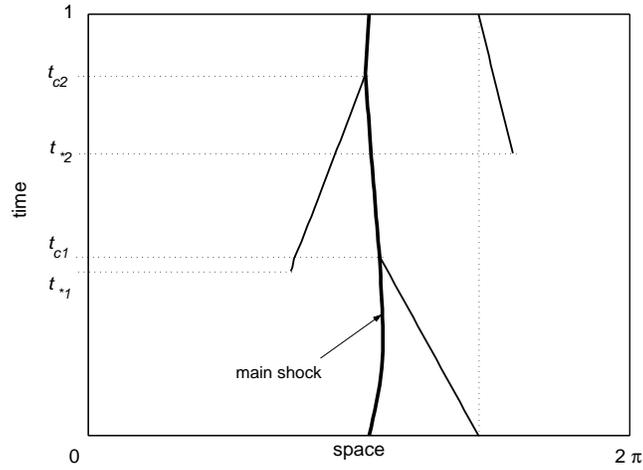,width=8.5cm}}
\else\drawing 65 10 {shock positions}
\fi
\vspace{2mm}
\caption{Evolution of shock positions during one period. The
beginnings of lines correspond to births of shocks (preshocks) at
times $t_{\star 1}$ and $t_{\star 2}$; shock mergers take place at
times $t_{c1}$ and $t_{c2}$. The ``main shock'', which survives for
all time, is shown with a thicker line.}
\label{f:shock-position}
\end{figure}

We find that, for all initial conditions $u_0(x)$ used, the solution
$u(x,t)$ relaxes exponentially in time to a unique function
$u_\infty(x,t)$ of period 1 in time.  Figure~\ref{f:relax-expo} shows
the variation of $\int_0^{2\pi}|u(x,n_-)-
u_\infty(x,1_-)|\,dx/(2\pi)$ for three  different initial conditions
as a function of the discrete time $n$.

The phenomenon of exponential convergence to a unique space- and
time-periodic solution is something quite general: whenever the
kicking potential $G(x)$ is periodic and analytic and the initial
velocity potential is periodic (so that the mean velocity $\la u\ra$
=0 at all times), there is exponential convergence to a unique
piecewise analytic solution. This is proved rigorously in the Appendix
for functions $G(x)$ which have a unique point of maximum with a
nonvanishing second derivative (Morse generic functions). Here, we
just explain the main ideas of the proof and give some additional
properties of the unique solution.

One very elementary property of solutions is that, for any initial
condition of zero mean value, the solution after at least one kick
satisfies $|u(x,t)|\le (1/2) + \max_x |dG(x)/dx|$. Indeed, at a time
$t=n_-$ just before any kick we have $x= y+ u(x,n_-)$ where $y$ is the
position just after the previous kick of the fluid particle which goes
to $x$ at time $n_-$. It follows from the spatial periodicity of the
velocity potential that the location $y$ which minimizes the action is
within less than half a period from $x$. Thus, $|u(x,n_-)|\le
1/2$. The additional $\max_x |dG(x)/dx|$ term comes from the maximum
change in velocity from one kick. It follows that the solution is
bounded. Note that if the spatial and temporal periods are $L$ and
$T$, respectively, the bound on the velocity becomes $L/(2T) + \max_x
|dG(x)/dx|$.

The convergence at large times to a unique solution is related to
properties of the two-dimensional conservative (area-preserving)
dynamical system defined by the Euler--Lagrange map
(\ref{map1})-(\ref{map2}) of \S\ref{s:minimizer}. By construction, we
have $u(x,1_+)=\hat{u}(x) -dG(x)/dx$, where $\hat{u}(x)$ is the
solution of the unforced Burgers equation at time $t=1_-$ from the
initial condition $u(x)$ at time $t=0_+$. The map $u\mapsto \hat{u}(x)
+g(x)$, where $g(x)\equiv -dG(x)/dx$, will be denoted $B_g$ and is just
the map which solves the kicked Burgers equation over a time of one.
The problem is to show that the iterates $B_g^{n}u_0$ converge for
$n\to \infty$ to a unique solution.

If it were not for the shocks it would suffice to consider 
the two-dimensional Euler--Lagrange map.
Note that, for the case of periodic kicking, this  map
has an obvious fixed point $P$, namely $(x=x_c,v=0)$, where $x_c$ is the
unique point maximizing the kicking potential.  It is easily
checked that this fixed point is an unstable (hyperbolic) saddle point of the
Euler--Lagrange map with two eigenvalues $\lambda = 1 + c +\sqrt{c^2 +
2c}$ and $1/\lambda$, where $c=-(1/2)(d^2G(x_c)/dx^2)$.

\begin{figure}
\iffigs 
\centerline{\psfig{file=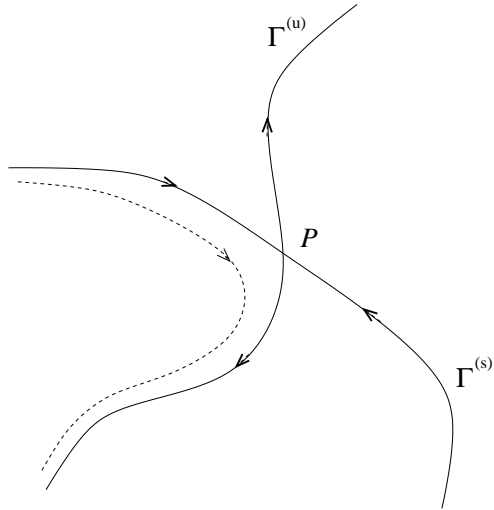,width=6.5cm}}
\else\drawing 65 10 {stable and unstable manifolds}
\fi
\vspace{2mm}
\caption{Sketch of a hyperbolic fixed point $P$ with stable ($\Gamma ^{\rm
(s)}$) and unstable ($\Gamma ^{\rm (u)}$) manifolds. The dashed line
gives the orbit of successive iterates of a point near the stable manifold.}
\label{f:stable-instable}
\end{figure}

Like for any two-dimensional map with a hyperbolic fixed point, there
are two curves globally invariant by the map which intersect at the
fixed point: the stable manifold $\Gamma ^{\rm (s)}$, the set of
points which converge to the fixed point under indefinite iteration of
the map, and the unstable manifold $\Gamma ^{\rm (u)}$, the set of
points which converge to the fixed point under indefinite iteration of
the inverse map, as illustrated in figure~\ref{f:stable-instable} (see,
e.g., H\'enon 1983, Manneville 1990). 
It follows that any curve which intersects the stable manifold
transversally (the tangents of the two curves are distinct) will,
after repeated applications of the map, be pushed exponentially
against the unstable manifold at a rate determined by the eigenvalue
$1/\lambda$.  In the language of Burgers dynamics, the curve in the
$(x,v)$ plane defined by an initial condition $u_0(x)$ will be mapped
after time $n$ into a curve very close to the unstable manifold. In
fact, for the case studied numerically, $1/\lambda\approx 0.18$ is
within one percent of the value measured from the exponential part of
the graph shown in figure~\ref{f:relax-expo}. Note that if the initial
condition $u_0(x)$ contains the fixed point, the convergence rate
becomes $\left(1/\lambda\right)^2$ (even higher powers of $1/\lambda$
are possible if the initial condition is tangent to the unstable
manifold).

\begin{figure}
\iffigs 
\centerline{\psfig{file=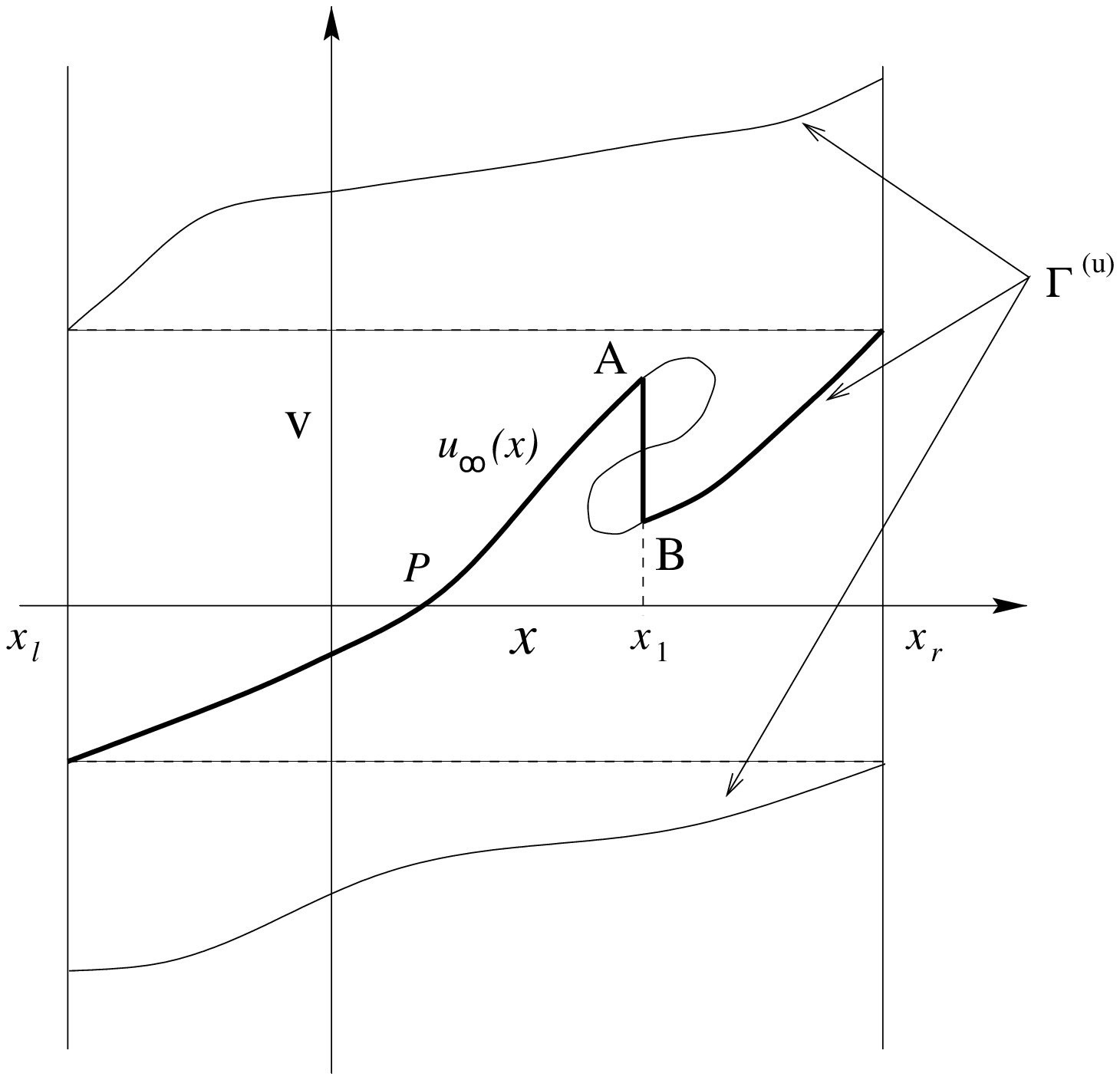,width=10cm}}
\else\drawing 65 10 {unstable manifold and Burgers}
\fi
\vspace{2mm}
\caption{Unstable manifold $\Gamma ^{\rm (u)}$ on the $(x,v)$-cylinder
(the $x$-coordinate is defined modulo~1) which passes through the
fixed point $P=(x_c,0)$. The bold line is the graph
of $u_\infty(x,1_-)$. The main shock is located at
$x_l=x_r$. Another shock at $x_1$ corresponds to a local zig-zag of
$\Gamma ^{\rm (u)}$ between A and B.}
\label{f:unstable}
\end{figure}

The fixed point $P$ gives rise to a very simple global minimizer:
$(y_j =x_c,\, v_j=0)$ for all positive and negative $j$'s. It follows
indeed by inspection of (\ref{actionz}) that any deviation from this
minimizer can only increase the action; actually, it minimizes both
the kinetic and the potential part of the action. Note that the
corresponding fluid particle is at rest forever and will never be
captured by a shock (it is actually the only particle with this
property).  It is shown in the Appendix that any minimizer is
attracted exponentially to such a global minimizer as $t\to
-\infty$. Thus, any point $(y_j,v_j)$ on a minimizer belongs to the
{\em unstable manifold\/} $\Gamma ^{\rm (u)}$ and, hence, any regular
part of the graph of the limiting solution $u_\infty(x)$ belongs to
the unstable manifold $\Gamma ^{\rm (u)}$. This unstable manifold is
analytic in the relevant region but can be quite complex. It can have
several branches for a given $x$ (see figure~\ref{f:unstable}) and
does not by itself define a single-valued function $u_\infty(x)$. The
solution has shocks and is only piecewise analytic. Consideration of
the minimizers is required to find the position of the shocks in the
limiting solution: two points with the same $x$ corresponding to a
shock, such as A and B on figure~\ref{f:unstable} should have the same
action.

\begin{figure}
\iffigs 
\centerline{\psfig{file=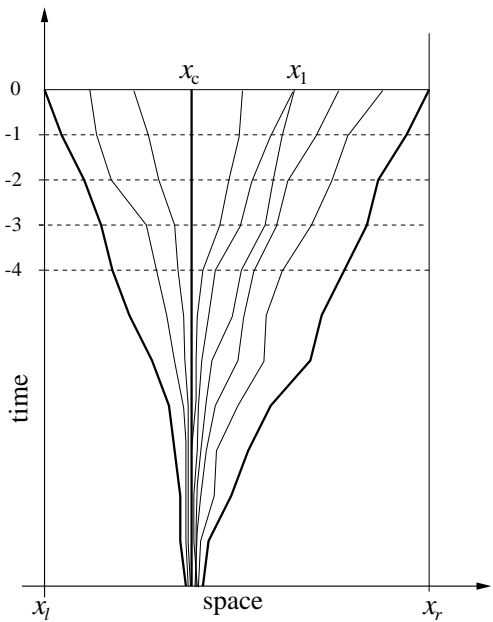,width=6.5cm}}
\else\drawing 65 10 {main shock}
\fi
\vspace{2mm}
\caption{Minimizers (trajectories of fluid particles) on the
$(x,t)$-cylinder. Time starts at $-\infty$. Shock locations at $t=0_-$
are characterized by  having two minimizers (an instance is at $x_1$).
The main shock is at $x_l=x_r$. The fat line $x=x_c$ is the global minimizer.}
\label{f:main}
\end{figure}

Finally, we give the geometric construction of the main shock, the
only shock which exists for an infinite time.  Since $\lambda$ is
positive, locally, minimizers which start to the right of $x_c$
approach the global minimizer from the right, and those which start to
the left approach it from the left. Take the rightmost and leftmost
points $x_r$ and $x_l$ on the periodicity circle such that the
corresponding minimizers approach the global minimizer from the right
and left respectively (see figure~\ref{f:main}). These points are
actually identical since there cannot be any gap between them that
would have minimizers approaching the global minimizer neither from
the right nor the left. The solution $u_\infty(x)$ has then its main
shock at $x_l=x_r$.

\section{Statistical properties for the periodically kicked case}
\label{s:statistical}

We are here working with time- and space-periodic deterministic
solutions of the kicked Burgers equation.  We thus choose to define
our statistical averages as averages over the two periods, here assumed
to be both unity. Specifically, let ${\cal F}(u)$ be an observable
(functional of the solution $u$) and let $T_{x,t}$ denote the
space-time translation operator which shifts the  solution $u$ by a
spatial amount $x$ and a temporal amount $t$.  We define
\begin{equation}
\la{\cal F}(u)\ra \equiv \int_0^1 \int_0^1{\cal F}
\left(T_{x,t}\,u\right)\,dx\,dt.
\label{defaverage}
\end{equation}
For example, with the observable ${\cal F}(u)\equiv \left[u(\Delta
x,0)-u(0,0)\right]^p$, we obtain the structure function of order $p$
over a separation $\Delta x$:
\begin{equation}
S_p(\Delta x) \equiv \int_0^1 \int_0^1 \left[u(x+\Delta x,t)-u(x,t)\right]^p
dx\, dt.
\label{defstructspactemp}
\end{equation}

Such averages are easily calculated numerically. For example, 
pdf's are obtained  from space-time histograms over all collocation
points and a suitably large number of output times.

\subsection{Pdf's of velocity derivatives and increments}
\label{s:pdf}

\begin{figure}
\iffigs 
\centerline{\psfig{file=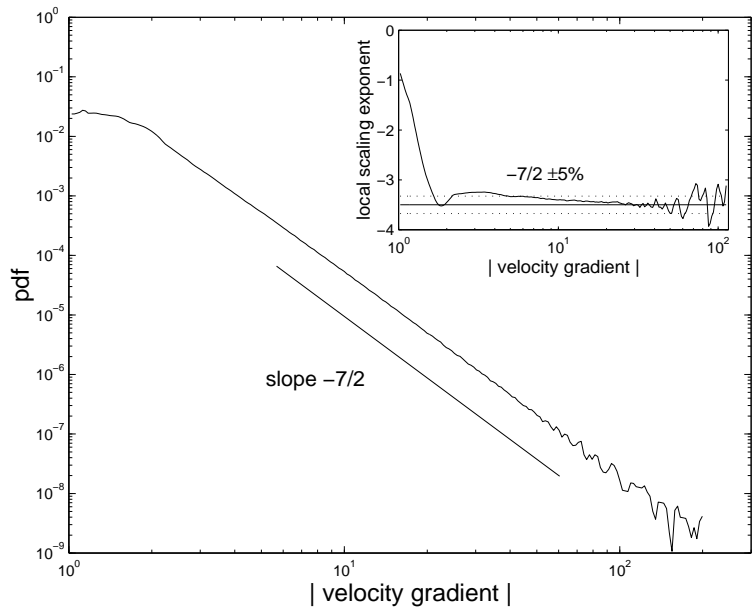,width=10cm}}
\else\drawing 65 10 {pdf gradient. log-log}
\fi
\vspace{2mm}
\caption{Pdf of the velocity gradient at negative values in log-log
coordinates. Upper inset: local scaling exponent. A power law with
exponent $-7/2$ is obtained at large arguments.}
\label{f:first-deriv}
\end{figure}

\begin{figure}
\iffigs 
\centerline{\psfig{file=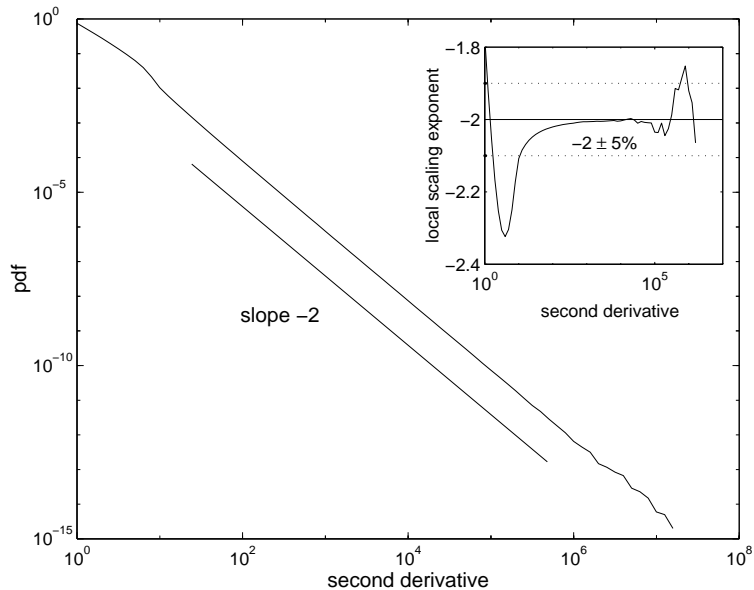,width=10cm}}
\else\drawing 90 10 {pdf second deriv. log-log. To be implemented}
\fi
\vspace{2mm}
\caption{Same as figure~\ref{f:first-deriv} with the second space
derivative of the velocity. The exponent is now $-2$.}
\label{f:second-deriv}
\end{figure}

For the periodic solution of \S\ref{s:periodic} we calculate first and
second space derivatives of the velocity; the corresponding pdf's are
then determined as normalized space-time histograms after binning of
derivative values (the bins are in geometric progression ; there are
100 bins per decade for the first derivative and 50 for the second).
Figures~\ref{f:first-deriv} and \ref{f:second-deriv} show the pdf's of
the first and second space derivatives in log-log
coordinates. Negative values are shown for the former and positive
values for the latter. It is seen that clean power laws are
obtained. More quantitative information about the values of the
exponents of the power laws are obtained by measuring the ``local
scaling exponent'', i.e.\ the logarithmic derivative of the pdf,
calculated here using least-square fits on quarter decades. The
results are shown as upper right insets on figures~\ref{f:first-deriv}
and \ref{f:second-deriv}. It is seen that, over one decade, the local
exponent for the pdf of the gradient is within less than five percent
of the value $-7/2$ predicted by a simple theoretical arguments given
hereafter; for the second derivative there are about four decades
within five percent of the value $-2$.

The presence of a power-law pdf with $-7/2$ exponent is easily
understood. It is just the signature of the preshocks which appear
when new shocks are created during the unforced phase between two
successive kicks. These preshocks are the only structures giving large
{\em finite\/} negative gradients: shocks give infinite negative
gradients (unless a finite viscosity is introduced) and the gradients in
the immediate spatial neighborhood of a mature shock are not
particularly large. The theory of the $-7/2$ law at large negative
values $\xi$ of the velocity gradient, developed by Bec \& Frisch
(2000) for the unforced case with {\em random\/} initial conditions is
readily adapted to the deterministic case, provided we use space-time
averages instead of ensemble averages. A simplified presentation,
following in part E {\it et al}.\ (1997), is given hereafter for the
case of a single preshock. The contributions of several preshocks to
the pdf are just additive and it is proved in the Appendix that the
periodic solution has finitely many preshocks.

We define a velocity in Lagrangian coordinates $u ^{\rm (L)}(a,t')$
with the origin of time just after a kick and $a=0$ at a (negative)
minimum of $\partial_x u$. Without loss of generality, we assume $u
^{\rm (L)}(0,0)=0$ (otherwise we perform a Galilean transformation to
bring it to zero). We then have, locally, $u ^{\rm (L)}(a,0)= -c_1a
+c_2 a ^3 +{\rm h.o.t.}$, where $c_1$ and $c_2$ are positive constants
and ``h.o.t.''  stands for higher-order terms. No generality is lost
by assuming $c_1=1$ (otherwise make a linear change on the
$a$-coordinate). The fluid particle initially at $a$ will be at time
$t'$ at $x=a+t'u ^{\rm (L)}(a,0)= a(1-t')+t'c_2 a ^3 +{\rm h.o.t.}$
This ``Lagrangian map'' becomes singular at $t'=t_\star= 1$, the
instant of preshock (formation of a shock).  We then have $a=
(x/c_2)^{1/3}+{\rm h.o.t.}$ (The cubic root is here defined both for
positive and negative values of its argument.)  Since the Lagrangian
velocity has not changed, the Eulerian velocity is given by
$u(x,t_\star)= - (x/c_2)^{1/3}+{\rm h.o.t.}$, which has a cubic root
structure and a gradient $ -(x/c_2)^{-2/3}/(3c_2)$. Hence, the
gradient takes large negative values for small $x$.  Just before
$t_\star$, at time $t'=1-\tau$, we have, $x=\tau a+c_2 a ^3 +{\rm
h.o.t.}$ It follows that the cubic relation between $a$ and $x$ still
holds, except in a region of Lagrangian width $\sim \tau ^{1/2}$ and
thus of Eulerian width $\sim \tau ^{3/2}$, where the relation becomes
linear to leading order.  (It is because of this $\tau ^{3/2}$
dependence that the time $\tau$ between successive outputs and the
mesh $\delta x =2\pi/N$ must be related by $\delta x \sim \tau
^{3/2}$.)

The question is now: what is the fraction of space-time where the
velocity gradient $\partial_x u < \xi$, where $\xi$ is a large
negative number\,? Because of the cubic root structure, $x$ must be in
a small interval of width $\sim |\xi|^{-3/2}$. The time must be
sufficiently close to $t_\star$ for this interval still to be in the
region of validity of the cubic relation, that is, within $\sim
|x|^{2/3} \sim |\xi|^{-1}$. Hence, the relevant space-time fraction
or, in other words, the cumulative probability to have $\partial_x u <
\xi$ is $\sim |\xi|^{-5/2}$. This gives a pdf $\sim |\xi|^{-7/2}$ at
large negative $\xi$'s.

Actually, there is another contribution, also proportional to 
$|\xi|^{-7/2}$  stemming from  a small time interval $\tau \sim
|x|^{2/3} \sim |\xi|^{-1}$ just {\em after\/} $t_\star$ when small-amplitude
shocks are present which have not yet completely destroyed the cubic
root structure (Bec \& Frisch 2000). For  the case studied
numerically, where the kicking potential is given by (\ref{defG}),
there are two  preshocks, each giving a contribution to the pdf of the
gradient proportional to $|\xi|^{-7/2}$.

This argument is readily adapted to second space derivatives, yielding
a pdf $\sim |\xi|^{-2}$ as observed in
figure~\ref{f:second-deriv}. (The same law holds also at large
positive values since the second derivative near a preshock is an even
function.)

\begin{figure}
\iffigs 
\centerline{\psfig{file=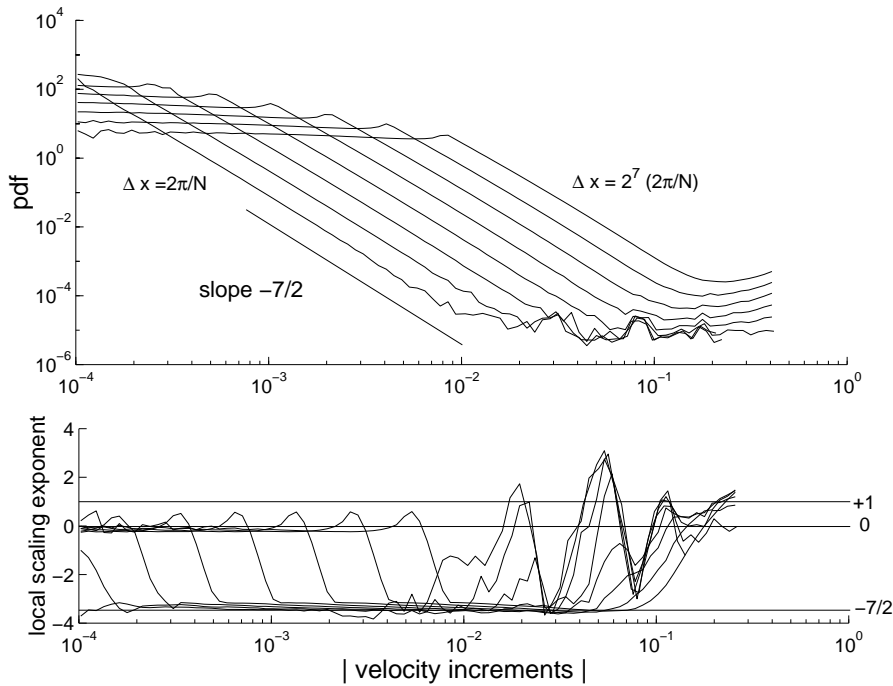,width=12cm}}
\else\drawing 65 10 {pdf increments}
\fi
\vspace{2mm}
\caption{Upper part: Pdf of (negative) velocity increments in log-log
coordinates for various values of the separation $\Delta x$ in
geometric progression from $2\pi/N$ to $2^7 (2\pi/N)$. Lower part: the
corresponding local scaling exponents.}
\label{f:increments}
\end{figure}

We now turn to the pdf of (spatial) velocity increments over
a separation $\Delta x$. We define
\begin{equation}
\Delta u(\Delta x;x,t)\equiv u(x+\Delta x,t)-u(x,t).
\label{defincrement}
\end{equation}
Its pdf's for various values of $\Delta x$ are again calculated from
space-time histograms. One hundred bins per decade are used.  $\Delta
x$ is given the values $2\pi 2 ^p/N$, where $2\pi /N$ is the numerical
mesh and $p$ is varied from zero to seven. Figure~\ref{f:increments}
gives log-log plots of the pdf's of increments for the eight
separations chosen. We limit ourselves to negative increments. The
corresponding local scaling exponents are shown in
figure~\ref{f:increments}.
It is seen that, for moderately large increments, the pdf's have the
same power-law behavior with exponent $-7/2$ as the pdf of the
gradient. This universal behavior was also predicted by E {\it et al}.\
(1997) for white-in-time forced Burgers
turbulence. Phenomenologically, this range is obtained simply by
Taylor expanding the increment as $\Delta x \,\partial_x u$. At larger
increments (in absolute value) the local scaling exponent rises
quickly to positive values but does not saturate to the value $+1$
predicted by E {\it et al}.\ (1997) by the following argument, based on the
consideration of nascent shocks and which applies also to the
periodically kicked case: The probability (as fraction of space) to
have a shock in an interval of length $\Delta x$ is $\propto\Delta
x$. Since the shock amplitude grows as $(t-t_\star)^{1/2}$, where
$t_\star$ is the time of the preshock, the fraction of time for which
the shock amplitude does not exceed a value $|\Delta u|$ is proportional
to $(\Delta u)^2$. Hence, the cumulative probability to have a
velocity increment (in absolute value) less than $|\Delta u|$ is
proportional to $\Delta x(\Delta u)^2$ and the pdf is proportional to
$\Delta x|\Delta u|$. By equating the contributions from the $-7/2$ and the
$+1$ ranges, the transition between the two ranges is predicted to
happen around an increment $\Delta u_c$  which scales as $(\Delta
x)^{1/3}$, in good agreement with our data. A clean $+1$ range is not
seen and  would require a resolution
of well over one million collocation points.

Finally, the flat range seen in figure~\ref{f:increments} for $|\Delta
u | \ll \Delta x$, is a universal contribution $\propto (\Delta
x)^{-1}$ from extremal points of the velocity (not predicted by E {\it
et al}.\ (1997)). This range extends also to positive  values of $\Delta
u \ll \Delta x$ but there is no other universal range for positive
increments; this is why pdf's are not shown for such increments.

\subsection{Structure functions}
\label{s:structure}

We now study the structure functions for the limiting (unique)
solution $u_\infty(x,t)$ of the periodically kicked Burgers equation.
For numerical studies $2\pi$-periodicity in space and 1-periodicity in
time are assumed. Hence, the
structure function of (integer) order $p$ is given by 
\begin{equation}
S_p(\Delta x) \equiv {1\over 2\pi}\int_0^1 dt\int_0^{2\pi}dx
\left[u_\infty(x+\Delta x,t) -u_\infty(x,t)\right]^p.
\label{defstruct}
\end{equation}
The $2\pi$-periodicity of $u_\infty(x,t)$ immediately implies that
$S_p(\Delta x)$ is  $2\pi$-periodic in $\Delta x$ and is an even/odd
function for even/odd  $p$.

\begin{figure}
\iffigs 
\centerline{\psfig{file=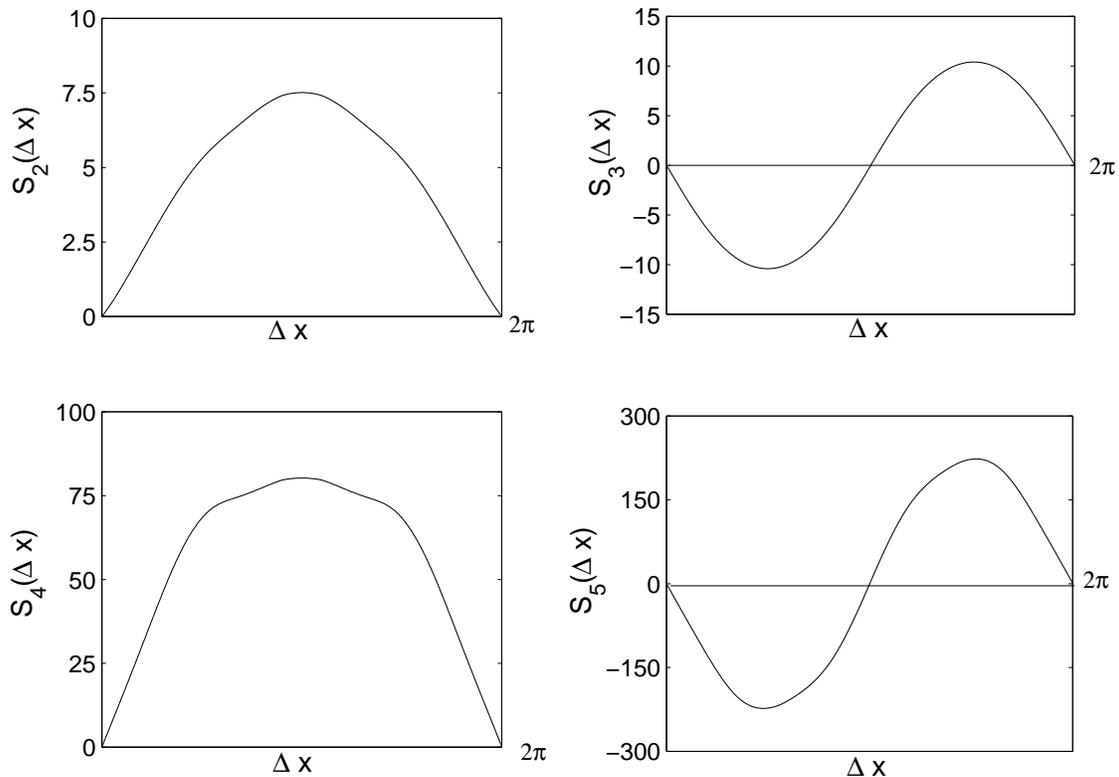,width=15cm}}
\else\drawing 65 10 {structure functions of order 2 and 3}
\fi
\vspace{2mm}
\caption{Structure functions $S_p(\Delta x)$ for $p=2,3,4,5$ as
labelled. Note the linear behavior at small $\Delta x$.}
\label{f:s2-5}
\end{figure}

Figure~\ref{f:s2-5} shows the structure functions of
order 2, 3, 4 and 5 (as labelled) in linear coordinates.  It is seen
that all these structure functions behave proportionally to $\Delta x$
at small arguments (more precisely as $\Delta x\left({\rm
sign}\,(\Delta x)\right)^{p+1}$). This is a well-known consequence of the
presence of shocks (E {\it et al}.\ 1997). 
In the next two sections we shall show that all the structure
functions except $S_3(\Delta x)$ are nonanalytic functions of $\Delta
x$.

\subsubsection{Analyticity of the third-order structure function}
\label{s:analytic}

For notational convenience in this and the next sections we assume
1-periodicity in space and time. Space averages over the period are
denoted $\la\cdot\ra_x$. Averages over both space and time are denoted
$\la\cdot\ra$. We shall prove that, when the kicking potential $G(x)$
is analytic, the third-order structure function is also analytic.

This will be established as a consequence of the following  relation 
for the unforced Burgers equation with space-periodic solution in the 
limit of vanishing viscosity. Let $u\equiv u(x,t)$ and $u'\equiv
u(x+\Delta x,t)$, we have
\begin{equation}
\partial_t \la u'u\ra_x = {1\over6}\partial_{\Delta x}\la (u'-u)^3\ra_x.
\label{12law}
\end{equation}
It is here assumed that $\Delta x$ is not an integer multiple of the
spatial period and that no pair of shocks remains separated by exactly
$\Delta x$ for a finite amount of time (this holds for almost every
$\Delta x$).

\smallskip
\par\noindent {\bf Proof.}\,\, Let us denote by $X_i(t)$
($i=1,\ldots,N(t)$) the (Eulerian) ordered positions of shocks and by
$\lsaut u \rsaut_i \equiv u(X_i(t)_+,t) - u(X_i(t)_-,t)$ the
(negative) velocity jump at the $i$-th shock. ($N(t)$ may change in
time.)  Except at shocks we can use the inviscid Burgers equation
$\partial_t u+u\partial_x u=0$. At shocks this has to be supplemented
by the Rankine--Hugoniot conditions (see, e.g. Lax 1957)
\begin{equation}
\dot X_i \lsaut u \rsaut_i - \lsaut {u ^2\over2} \rsaut_i =0,
\label{sautises}
\end{equation}
which follow also from  momentum conservation applied to small intervals
straddling the shocks. An immediate consequence is that the velocity
of the shocks are given by
\begin{equation}
\dot X_i(t) = {1\over 2}\left[u(X_i(t)_+,t) + u(X_i(t)_-,t)\right].
\label{halfsum}
\end{equation}
We observe that
\begin{equation}
\la u'u  \ra_x = \sum_{i=1}^{N}\int_{X_i(t)}^{X_{i+1}(t)}
u'u\,dx,
\label{eq:correlspat}
\end{equation}
where $X_{N+1}(t)\equiv X_1(t)+1$. Differentiating with respect to
$t$, we obtain
\begin{eqnarray}
\partial_t  \la u'u  \ra_x &=& \!\!\la u
\partial_t u'+u'\partial_t u  \ra_x - \sum_{i=1}^{N}
\dot{X}_i(t)\lsaut u \rsaut_i\, u(X_i(t)+\Delta x,t)\label{eq:dertspaccor}\\
&=&\!\!\!-\sum_{i=1}^{N}
\int_{X_i(t)}^{X_{i+1}(t)}\!\!\left(uu'\partial_x u'+u'u\partial_xu
\right)dx - \!\sum_{i=1}^{N} \dot{X}_i(t)\lsaut u
\rsaut_i\, u(X_i(t)+\Delta x,t)
\label{eq:dertspaccorencor}\\
 &=&\!\!\!-\sum_{i=1}^{N}
\int_{X_i(t)}^{X_{i+1}(t)}\left(\frac{1}{2}u\partial_{\Delta x}u'^2-
\frac{1}{2}u^2\partial_{\Delta x} u'\right)dx \nonumber \\ &&\!\!\!
- \sum_{i=1}^{N} \dot{X}_i(t)\lsaut u \rsaut_i\,u(X_i(t)+\Delta x,t)
+\sum_{i=1}^{N} \lsaut {u^2\over 2} \rsaut_i\,u(X_i(t)+\Delta x,t).
\label{eq:avantder}
\end{eqnarray}
In going from (\ref{eq:dertspaccor}) to (\ref{eq:dertspaccorencor}) we
used the inviscid decaying Burgers equation; from (\ref{eq:dertspaccorencor})
to (\ref{eq:avantder}) we have performed an integration by parts 
and used $\partial_x u'=\partial_{\Delta x}u'$. From (\ref{sautises})
follows that  the last two terms in (\ref{eq:avantder})
cancel. Hence, we obtain
\begin{equation}
\partial_t  \la u'u  \ra_x ={1\over2}\partial_{\Delta x} 
 \la -uu'^2+u^2u'
\ra_x
 = {1\over6}\partial_{\Delta x}\la (u'-u)^3\ra_x,
\label{onyest}
\end{equation}
which completes the proof.

We now return to the case of the periodically kicked Burgers equation,
with the unique solution $u_\infty(x,t)$. Using (\ref{12law}),
integrated in time between two successive kicks, say at $t=0$ and
$t=1$, we have
\begin{eqnarray}
\!\!\!\!\!\!{1\over6}\partial_{\Delta x}S_3(\Delta x) &=&
{1\over6}\partial_{\Delta x}\la \left [u_\infty(x+\Delta
x,t)-u_\infty(x,t)\right]^3\ra \nonumber \\
&=&\la u_\infty(x+\Delta x,1_-)u_\infty(x,1_-)\ra_x -
\la u_\infty(x+\Delta x,0_+)u_\infty(x,0_+)\ra_x.
\label{eq:inttempkkh}
\end{eqnarray}
Next, we use
\begin{equation}
u_\infty(x,0_+)= u_\infty(x,0_-)+g(x)=u_\infty(x,1_-) +g(x),
\label{discont}
\end{equation}
which follows from (\ref{discon}) and (\ref{forceperiodic}); here,
$g(x)=-dG(x)/dx$ where $G(x)$ is
the kicking potential. Substituting this in (\ref{eq:inttempkkh}), we
obtain
\begin{equation}
{1\over6}\partial_{\Delta x}S_3(\Delta x)=\la g(x)
g(x+\Delta x)\ra_x-\la g(x)u_\infty(x+\Delta
x,0_+)\ra_x - \la g(x+\Delta x)u_\infty(x,0_+)\ra_x.
\label{preanalyticity}
\end{equation}
We now assume that the kicking potential and, hence, $g(x)$ are 
analytic functions  and we find that all three terms on the r.h.s.\ of
(\ref{preanalyticity}) are analytic functions of $\Delta x$. This
follows indeed from the observation that the analyticity of $g(x)$ and
the boundedness of $h(x)$ imply the analyticity in $\Delta x$ of the
integral $\int_0^1 g(x)h(x+\Delta x)\,dx$, which is basically a
convolution integral. We have thus proved the analyticity of the third-order
structure function in the separation $\Delta x$.

When the kicking potential $G(x)$ has only a finite number of Fourier
harmonics a stronger result holds: the third-order structure function
has exactly the same harmonics as the kicking potential. This follows
because the r.h.s. of (\ref{preanalyticity}) is a convolution
integral. For the case of the kicking potential given by (\ref{defG}),
which has the harmonics of wavenumber 1 and 3, we thus have
\begin{equation}
S_3(\Delta x) = \lambda \sin (\Delta x) + \mu \sin (3\Delta x).
\label{lambdamu}
\end{equation}
(The presence of only sine functions is due to the odd character of
the third-order structure function.)  We have indeed checked that the
structure function $S_3(\Delta x)$ calculated numerically at the
beginning of \S\ref{s:structure} has a global fit of this form with
$\lambda \approx -10.9953$ and $\mu\approx -1.1463$ with an error of
less than $10^{-5}$.

We finally observe that the analyticity result for the third-order
structure function is quite general and has been proved also for the
case of white-noise forcing (E \& Vanden Eijnden 2000).

\subsubsection{Nonanalyticity of the structure functions of order $p\neq 3$}
\label{s:nonanalytic}

We now concentrate on integer values of $p>1$. Indeed, for non-integer values,
the structure function is not defined, unless we take the absolute
value of the velocity increment which results trivially in
nonanalyticity and, for $p=1$, the structure function vanishes.
We intend to show that
\begin{equation}
S_p(\Delta x)=\cases{A_p|\Delta x| + B_p (\Delta x)^2 
+o\left((\Delta x)^2\right),&for even $p$;\cr
A_p\Delta x + B_p \Delta x|\Delta x|
+o\left((\Delta x)^2\right), &for odd  $p$,\cr}
\label{structurecases}
\end{equation}
where the constant $A_p$ never vanishes and the constant $B_p$
vanishes for $p=3$ and never vanishes for $p>3$ (the expressions of
these constants will be given below). This will then imply (i) that
all structure functions are proportional to the first power of the
separation (a well-known result; see, e.g.\ E {\it et al}.\ 1997) and
(ii) that all structure functions of order $p\neq 3$ are nonanalytic
functions of $\Delta x$. Actually, we shall establish
(\ref{structurecases})
only for $\Delta x>0$; the extension to $\Delta x<0$ follows
then from the even/odd character of structure functions of even/odd orders.

The idea of the proof is to observe that the  only possible
sources of nonanalyticity  are singularities of the
solution in the space-time domain, namely, preshocks,  shocks
and shock mergers. The contributions from the analytic regions to
$S_p(\Delta x)$ is clearly $O\left((\Delta x)^p\right)$ and must
therefore be retained in (\ref{structurecases}) only for $p=2$.
Let us now concentrate on the contributions from singularities.

It is easily shown that preshocks  contribute at most
terms $O\left((\Delta x)^{(p+5)/3}\right)$ which are only higher order
corrections to (\ref{structurecases}). This follows from the scaling
properties of the pdf of increments as discussed in \S\ref{s:pdf} and
in Bec \& Frisch (1999) which is itself a consequence of the cubic
root structure of preshocks.

As to the contribution of (mature) shocks, we obtain it by first
calculating the contribution, denoted $S^{\rm shock}_p(\Delta x)$,
coming from the neighborhood of individual shocks, ignoring shock
mergers; then we determine the correction due to mergers,  denoted 
$\Delta S^{\rm merge}_p(\Delta x)$.

Let $X_j(t)$ denote the positions of the various shocks (their number
may change in time). Let $u ^+_j(x,t)$ and $u ^-_j(x,t)$ denote the
velocity in the immediate right and left neighborhood of the $j$-th
shock. Let $C_j(t)\equiv u_j^+(X_j(t),t) - u_j^-(X_j(t),t)$ denote the
(negative) jump at $X_j(t)$. (The jump $C_j(t)$ is taken equal to zero when a
shock has not yet been born or has disappeared by merger.) Since we
take $\Delta x>0$ the requirement that $x$ and $x+\Delta x$ straddle
the $j$-th shock limits the domain of $x$-integration to the interval
$]X_j(t)-\Delta x,X_j(t)[$. Hence, we have
\begin{equation}
S_p^{\rm shock}(\Delta x) = \int_0^1dt\sum_j \int_{X_j(t)-\Delta x}^{X_j(t)}dx
\left[u^+_j(x+\Delta x,t) - u^-_j(x,t) \right ]^p.
\label{eq:contribI}
\end{equation}
Since $u ^+_j(x,t)$ and $u ^-_j(x,t)$ are smooth functions of $x$, we
can  Taylor expand them near $X_j(t)$. For our interest
only the first two terms are relevant. We thus obtain
\begin{eqnarray}
S_p^{\rm shock}(\Delta x) = \!\!\!\!\!&
\displaystyle\int_0^1dt \sum_j  \int_{X_j(t)-\Delta
x}^{X_j(t)}dx \Biggl \{ u_j^+\left(X_j(t),t\right)
+\left[\partial_x u_j^+\left(X_j(t),t\right)
\right] (x+\Delta x-X_j(t)) \nonumber \\ &
-u_j^-\left(X_j(t),t\right)-\left[\partial_x
u_j^-\left(X_j(t),t\right) 
\right] (x-X_j(t))
\Biggr \}^p + o\left((\Delta x)^2\right).
\label{eq:approxSI}
\end{eqnarray}
We then use the following relation which governs the evolution of 
shock jumps:
\begin{equation}
\frac{d}{dt}C_j(t) = - \left[\partial_x u_j^+(X_j(t),t)+\partial_x
u_j^-(X_j(t),t)\right]\frac{C_j(t)}{2}.
\label{eq:relCdxu}
\end{equation}
(This relation is obtained by using the inviscid Burgers equation
on both sides of the shock and (\ref{halfsum}).) Using (\ref{eq:relCdxu})
in (\ref{eq:approxSI}), performing all the space integrals and keeping
only terms up to  $O\left((\Delta x)^2\right)$, we obtain
\begin{equation}
S_p^{\rm shock}(\Delta x) = \int_0^1dt \sum_j \left [  C_j^p(t) \Delta x -
\frac{p}{p-1} \frac{d}{dt}\left(C_j^{p-1}(t)\right)(\Delta x)^2\right
] + o\left((\Delta x)^2\right) .
\label{eq:finalI}
\end{equation}

\begin{figure}[h]
\iffigs 
\centerline{\psfig{file=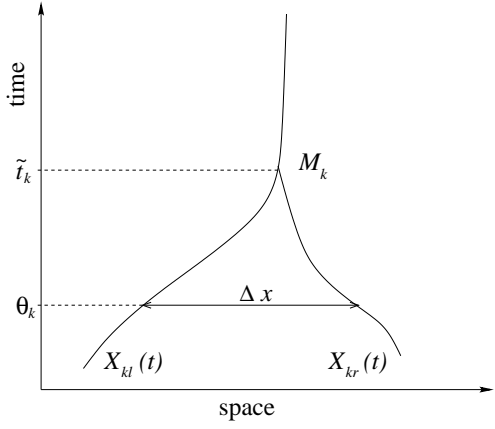,width=6.5cm}}
\else\drawing 65 10 {merger}
\fi
\caption{Merger of two shocks $X_{kl}(t)$ and $X_{kr}(t)$ on the
$(x,t)$-cylinder. The merger (event $M_k$)  takes place at $t=\tilde t_k$.
At the earlier time $t=\theta_k$ the shocks are within a distance
$\Delta x$.}
\label{f:merger}
\end{figure}
We turn to the contributions of the finite set $\{M_k\}$ of shock
mergers taking place in the periodic space-time domain. (We only
consider mergers of two shocks since events with more than two
shocks merging are not generic; furthermore, it may be checked that
they do not change our conclusions.) Associated
with each  event $M_k$, we define: $\tilde t_k$, the time of merger,
$X_{kl}(t)$ and $X_{kr}(t)$, the positions of the left and the
right shocks about to merge and  $C_{kl}(t)$ and $C_{kr}(t)$, the
respective jumps across these merging shocks. Let $\theta_k$ be the
first instant of time $t$ when the distance $X_{kr}(t)-X_{kl}(t)$
becomes less than $\Delta x$. It is clear that, for small $\Delta x$,
we have $\tilde t_k - \theta_k =O(\Delta x)$ (see
figure~\ref{f:merger}). Furthermore, we have
\begin{equation}
X_{kr}(t)- X_{kl}(t)= \Delta x +(t-\theta_k){C_{kl}(\tilde{t}_k)+
C_{kr}(\tilde{t}_k) \over 2}+o(\Delta x).
\label{mouvmtchocs}
\end{equation}
After $\theta_k$, an interval $]x,x+\Delta x[$ may either straddle a
single of the two merging shocks or both.  The calculation above did
not take into account the possibility of straddling two shocks. (For
$\theta_k<t<\tilde t_k$, this happens when $X_{kr}(t)-\Delta
x<x<X_{kl}(t)$.)  It is also necessary to modify the spatial
integration domain associated to situations where the interval of
length $\Delta x$ straddles a single shock. All these effects together
give a correction to (\ref{eq:finalI}) stemming from mergers which, to
leading order, reads
\begin{equation}
\Delta S_p^{\rm merge}(\Delta x)=-(
\Delta x)^2\sum_k{\left(C_{kl}(\tilde{t}_k)+C_{kr}(\tilde{t}_k)\right)^p -
C^p_{kl}(\tilde{t}_k)-C^p_{kr}(\tilde{t}_k)\over
C_{kl}(\tilde{t}_k)+ C_{kr}(\tilde{t}_k)}+o\left((\Delta x)^2\right).
\label{correctionend}
\end{equation}

We now assemble the various contributions. The second term in the
integral on the r.h.s. of (\ref{eq:finalI}) can be integrated
explicitly as a sum of terms coming from the birth and death of
shocks.  Preshocks do not contribute because they have vanishing
jumps. Shock mergers give three contributions: two from the incoming
shocks with jumps $C_{kl}(\tilde t_k)$ and $C_{kr}(\tilde t_k)$
and one from the merged shock with jump $C_{kl}(\tilde
t_k)+C_{kr}(\tilde t_k)$. We finally obtain the following expressions
for the coefficients in the expansion (\ref{structurecases}) of the
structure functions in terms of the shock jumps $C_j(t)$:
\begin{eqnarray}
A_p &=& \sum_j \int_0^1 C^p_j(t)\,dt,\quad \hbox{for all $p>1$}
\label{exprap} \\  
B_p &=&-\sum_k\Biggl \{ \frac{p}{p-1} \left[C^{p-1}_{kl}(\tilde t_k)
+C^{p-1}_{kr}(\tilde t_k)-\left(C_{kl}(\tilde t_k) +C_{kr}(\tilde
t_k)\right)^{p-1}\right]  \nonumber \\ 
&&+{\left(C_{kl}(\tilde{t}_k)+C_{kr}(\tilde{t}_k)\right)^p -
C^p_{kl}(\tilde{t}_k)-C^p_{kr}(\tilde{t}_k)\over
C_{kl}(\tilde{t}_k)+ C_{kr}(\tilde{t}_k)} \Biggr \}, \quad
\hbox{for all 
$p>2$}.
\label{exprbp}
\end{eqnarray}
For $p=2$, we must add to the expression given by the r.h.s. of
(\ref{exprbp}) the contribution from the analytic regions which
give also terms $O\left((\Delta x)^2\right)$. It is readily
seen that $B_p$ given by (\ref{exprbp}) vanishes for $p=3$ and only
for that value.  

It must be stressed that, for structure functions of order $p>3$, the
$\propto (\Delta x)^2$ corrections to the leading $\propto \Delta x$
terms come entirely from mergers. For cases which have a single shock
with no mergers and no preshocks (an instance is $G(x)=-\cos x$) this
correction is absent.

\section{Concluding remarks}
\label{s:conclusion}

We have shown that Burgers turbulence in the inviscid limit with
periodic large-scale kicking is characterized by universal properties
originally conjectured to hold for the case of random forcing.  In
particular, there is a power-law tail with exponent $-7/2$ in the pdf
of negative velocity gradients. This law was proposed by E {\it et
al}.\ (1997) for the case of random forcing which is smooth in space
and white noise in time. The validity of the $-7/2$ law for the latter
case is still an open question.  It is clear, that the $-7/2$ law is
unescapable as soon as preshocks are present and well separated. Some
hypothetical  clustering of preshocks could invalidate the
$-7/2$ law for the white-in-time case. Careful numerical
experimentation using a sequence of random forces which approach
white-in-time forcing should be able to shed some light on this
issue. This can in principle be done using kicked Burgers turbulence
in which the instants of kicking are taken closer and closer and the
successive spatial impulses are taken random and
independent. Obtaining sufficiently clean power-law scaling to
distinguish, e.g.\ between an exponent $-3$ and $-7/2$ in the random
case may require very large computational resources. It is also useful
to  investigate the statistical distribution of
preshocks to test their possible clustering properties.

Let us briefly now address the question of the effect of a finite
small viscosity. Basically, this will broaden the shocks giving them a
hyperbolic tangent structure of width $\propto \nu$. From this it is easily
inferred that the maximum negative gradient is $|\xi|_{\rm max}=O(\nu
^{-1})$ and that the shoulders of such viscous shocks contribute a
term $\propto \nu |\xi|^{-1}$ to the pdf of (negative) gradients
(Gotoh \& Kraichnan 1998). This term will dominate over the inviscid
contribution $\propto |\xi|^{-7/2}$ beyond a crossover value of the
gradient $|\xi|_c \propto \nu ^{-2/5}$. A small viscosity will also
regularize preshocks, giving them a finite velocity gradient
$|\xi|_{\rm max\,pr}\propto \nu ^{-1/2}$ (Crighton \& Scott
1979). This gives only subdominant contributions for all 
$\xi$'s.

The Burgers equation constitutes a dissipative dynamical system because of
the presence of shocks which introduce an essentially irreversible
element into the dynamics. Many features of our periodically kicked
Burgers problem are actually in exact correspondence with those of a
conservative (Hamiltonian) dynamical system, namely the equilibrium 
positions of a one-dimensional
chain of (classical) atoms connected by elastic springs in the
presence of a space-periodic external potential (Frenkel--Kantorova
model). This problem has been investigated by  Aubry (1983) and Mather
(1982). The potential energy which has to be minimized to obtain
the ground state has the following form:
\begin{equation}
H(\{y_j\}) = \sum_{j} {1\over2} (y_{j+1} - y_j -a)^2 - \varepsilon G(x),
\label{hamil}
\end{equation}
where the $y_j$'s are the positions of the atoms, $a$ is the
unstretched length of the springs and $\varepsilon > 0$ measures the
depth of the periodic external potential $\varepsilon G(x)$. Nontrivial
properties of the ground states reflect the competition between the
tendency of the atoms to sit at the minimum of the potential $-
\varepsilon G(x)$ and to be within a distance $a$ from each other.  It
is easily checked that the action (in the sense of \S\ref{s:kicked})
for the kicked Burgers equation with periodic forcing is exactly given
by the Aubry--Mather Hamiltonian (\ref{hamil}) if we take a forcing
potential $\varepsilon G(x)$ and a mean velocity $\la u\ra =a$. The
velocity in the Burgers equation is now the analog of the distance
between adjacent atoms. Note that $j$ is a space index in the chain
model, whereas it is a time index in the Burgers equation. In this
paper we have assumed a vanishing mean velocity ($a=0$).  In this case
there is no competition between the aforementioned two tendencies, and
the global minimizer $y_j = x_c$ corresponds to a trivial ground state
minimizing both parts of the Hamiltonian. More delicate effects
arising in the case $ \la u \ra \neq 0$ and in more than one dimension
will be discussed in forthcoming work.  We note that the connection between
Aubry--Mather theory and Burgers equation for the case of
time-periodic potentials with continuous time was discussed for the
first time in Jauslin, Kreiss \& Moser (1997) and in 
E {\it et al}.\ (2000). The theory has been recently developed further
in E (1999) and in Sobolevski (1999).

Finally, we discuss some possible extensions of the ideas of the
present paper to Navier--Stokes turbulence. Obviously, the method of
forcing by kicks applied at discrete instants of time can also be used
for Navier--Stokes. One of the things which made this method
particularly valuable for the Burgers case is the existence of an
efficient numerical algorithm to solve the purely decaying Burgers
equation in the inviscid limit. At the moment there is nothing
comparable for Navier--Stokes. Note that periodic kicking will not
result in a unique time-periodic solution for Navier--Stokes since
periodicity-breaking bifurcations leading to chaos will unavoidably occur.

The results concerning analyticity of structure functions are likely
to be the same for Burgers turbulence and isotropic 3-D Navier--Stokes
turbulence (in the limit of vanishing viscosity). Indeed, in the
latter case all structure functions of order $p\neq 3$ are generally believed
to have scaling properties at small separations with nontrivial (and
certainly noninteger) exponents. As to the third-order structure
function, our proof of analyticity for analytic-in-space forcing 
can be extended if one assumes that Kolmogorov's four-fifths law
is valid (Kolmogorov 1941; see also Chapter 6 of Frisch 1995).

Burgers turbulence presents an algebraic tail for the pdf of the
velocity gradient but nothing similar is known for Navier--Stokes
turbulence. This may be telling us something about the possible
singularities associated to the Navier--Stokes and Euler equations.  A
remarkable feature for the Burgers equation is that only preshocks but
not mature shocks contribute to the power-law tail. There is indeed a
basic difference between the two types of singularities.  For analytic
forcing, near a mature shock, the solution is not only piecewise
analytic (i.e.\ on each side of the shock) but uniformly so: the
radius of convergence of the Taylor series remains finite as one
approaches a mature shock. In contrast, when approaching a cubic root
preshock singularity, the radius of convergence goes to zero and
gradients become very large. It is the algebraic behavior of their
strength which causes the power-law tail. If the Navier--Stokes
equation, in the limit of vanishing viscosity, were to develop any
singularity of this kind (accompanied by algebraically large
gradients) it should also display a power-law-tail pdf. Further
increases in the quality of experimental and numerical turbulence or
convection data are needed to find if such singularities are really
ruled out.

\vspace{2mm}
\par\noindent {\bf Acknowledgments}

We are grateful to M.~Blank, W.~E, R.~Kraichnan, A.~Noullez, M.R.~Rahimi~Tabar,
Ya.~Sinai, E.~Vanden~Eijnden and B.~Villone for useful remarks.
The work of KK was supported by the  Leverhulme Trust 
(research grant RF\&G/10906).
Simulations were performed in the framework of the SIVAM project of
the Observatoire de la C\^ote d'Azur, funded by CNRS and MENRT. Part
of this work was done while one of the authors (UF) was attending the
1999 TAO Study Center, supported by ESF.

\section*{APPENDIX}
\appendix
\renewcommand{\theequation}{A.\arabic{equation}}
\setcounter{equation}{0}

\subsection*{Statement of the results for periodic kicking}

Here, we formulate and prove formally the statements presented
somewhat heuristically in
\S\ref{s:periodic}.  The kicking is assumed 1-periodic in both space
and time.  The force in the Burgers equation is given by
\begin{equation}
f(x,t) = -{d\over dx} G(x) \sum_{j=-\infty}^{+\infty} \delta(t-j),
\label{forceperiodic1}
\end{equation}
where $G(x)$ is a deterministic function of $x$ which is 1-periodic
and three times continuously differentiable ($G \in C^3$). For some of
the statements below it is assumed that $G$ is analytic.  We shall
also assume that the kicking potential $G(x)$ is generic in the Morse
sense. This implies that $G(x)$, considered on the circle $0 \leq x <
1$, has a unique point of maximum $x_c$ and that $G(x)$ is
non-degenerate at $x_c$, i.e.\ $d^2G(x_c)/dx^2 < 0$. Without loss of
generality we can assume that $G(x_c) = 0$. We denote by $c =
-(1/2)d^2G(x_c)/dx^2 > 0$.  The initial potential $\psi_{0}(x)$ is
also assumed 1-periodic. This implies that $\int^{1}_{0}u_0(x)dx = 0$,
where $u_0(x)=u(x,0_+)$ is the initial velocity.

We now solve the unforced Burgers equation between the times $t=0$ and $t=1$
and get $\hat{u}(x)\equiv u(x,1_-)$. Then, we determine
$u(x,1_+)=\hat{u}(x) + g(x)$, where $g(x) = - dG(x)/dx$. Denote
by $B_g$ the transformation from $u(x,0_+)$ to $u(x,1_+) :
B_{g}u = \hat{u} + g$. Clearly, $B_g$ transforms $L^1_0 = \{u(x) \in
L^1 [0,1],\, \int^{1}_{0}u(x)dx = 0\}$ into itself. Then, the following
statements hold.
\begin{enumerate}
\item[S1] The functional transformation $B_g$ has a unique fixed point
$u_{\infty}$: \ \ $B_{g}u_{\infty} = u_{\infty}$, \ \
$\max_{x}|u_{\infty}(x)| \leq 1/2 + \max_{x}|g(x)|$. The fixed point
$u_{\infty}$ is a function of bounded variation, it is continuous
everywhere except at the set of shock points, which is at most
countably infinite.

\item[S2] Let $x$ be a point of continuity for $u_{\infty}$. Then, for all
$u_0$ such that $\int^{1}_{0}u_0(x)dx = 0$, \ \ \
\begin{equation}
B_{g}^{n}u_0(x) \to u_{\infty}(x) \ \ {\rm as} \ \ n \to \infty. 
\label{convergence}
\end{equation}
 \item[S3] The unique global minimizer $\gamma_c$ corresponds to a particle
with zero velocity sitting at the point $x_c$ of maximum  kicking potential.
\item[S4] There exists a unique entropy weak solution\footnote{Entropy
solutions are the limit of viscous solutions as $\nu\to
0$ (Lax 1957).} $u_{\infty}(x,t)$ to the kicked Burgers equation in the
semi-infinite domain $]-\infty,T]$ with zero mean velocity. This
solution is 1-periodic in time and it is generated by $u_{\infty}(x)$:
\ \ $u_{\infty}(x,k+) = u_{\infty}(x)$ for all integer $k$. The
solution $u_{\infty}(x,t)$ satisfies the estimate:
$\max_{x,t}|u_{\infty}(x,t)| \leq 1/2 + \max_{x}|g(x)|$.
\item[S5] For arbitrary $t$ there exists a unique main shock.
\item[S6] Convergence in (\ref{convergence}) is exponentially fast in
$n$.  If $u_{\infty}$ is differentiable at $x$, then there
exists a constant $C(x)$ which does not depend on $u_0$
such that
\begin{equation} 
|B_{g}^{n}u_0(x) - u_{\infty}(x)| \leq C(x)\lambda^{-n},
\label{estimate1}
\end{equation}
where  $\lambda = 1+c+\sqrt{c^2+2c}>1$. If $x_c$ is a point of continuity for $u_0$ and $u_0(x_c) \neq 0$ then
there exists a constant $c(u_0) > 0$ such that for all $x$ the
following estimate holds:
\begin{equation}
|B_{g}^{n}u_0(x) - u_{\infty}(x)| \geq c(u_0)\lambda^{-n}.
\label{estimate2}
\end{equation}
\item[S7] If the kicking potential $G(x)$ is analytic, then for any
$t$ the solution $u_{\infty}(x,t)$ is a piecewise analytic function of
$x$. The number of pieces is finite and is equal to the total number of
shocks at time $t$. The number of preshock events between time 0 and
1 is also finite.
\end{enumerate}

\subsection*{Proof} 

\subsubsection*{Uniqueness of the solution}

Denote by ${\cal T}$ the Euler--Lagrange diffeomorphism of the phase-space
cylinder ${\cal C}\equiv\{0 \leq y <1, -\infty < v < \infty\}$ generated by the
system of equations (\ref{map2}), i.e.\ ${\cal T}(y,v) = (y',v')$,
where
\begin{eqnarray}
&&v' = v + g(y)\\
&&y' = y + v + g(y)\ \ \  {\rm (mod\,\,1)}. \nonumber
\label{systemy1}
\end{eqnarray}
Since $G \in C^3$, the diffeomorphism $ {\cal T}\in C^2$. The global
minimizer in the sense of \S\ref{s:minimizer} is a sequence $\{ y_j =
x_c \}$ for all $j$. Here, it is trivial and it corresponds to a fixed
point $P = (x_c,0)$ for the diffeomorphism ${\cal T}$, i.e.\ the
corresponding trajectory of ${\cal T}$ is a stationary trajectory
$(y_j,v_j) = (x_c,0) = P$. Let $\{ y'_j, \ j \leq J, \ y'_J = x \}$ be
an arbitrary minimizer.  Then $y'_j \to x_c$ as $j \to
-\infty$. Indeed, if it were not the case, then a sequence $\{y''_J =
x,\ y''_j = x_c , \ j_0 \leq j \leq J - 1, \ y''_j = y'_j, \ j \leq
j_0 - 1 \}$ would have smaller action than $\{ y'_j \}$ for any $j_0$
which is sufficiently negative. (The action is given by
(\ref{actionz}).) Convergence $y'_j \to x_c$ implies that the
corresponding trajectory of ${\cal T}$, $\{ (y'_j, v'_j = \pm
\rho(y'_j,y'_{j-1})) \}$ converges to $P$ as $j \to -\infty$. Here,
$\rho$ denotes the distance between two points on the circle. Easy
calculation shows that $P$ is a saddle point for the diffeomorphism
${\cal T}$ with the eigenvalues $\lambda_1 = \lambda = 1 + c +\sqrt{c^2 + 2c}
> 1$ and $0 < \lambda_2 = 1/\lambda  < 1$. It follows that
there exist two $C^2$-smooth curves $\Gamma^{\rm (s)}$ and
$\Gamma^{\rm (u)}$ which are stable and unstable manifolds for the
point $P$. Both curves pass through $P$ and consist of all points
$(y,v)$ whose trajectories approach $P$ as $j$ tends to $+\infty$ and
$-\infty$, respectively (see figure~\ref{f:stable-instable}). In fact,
convergence is exponentially fast and the rate is given by
$1/\lambda$. Since $(y'_j, v'_j) \to P$ as $j \to -\infty$, a point
$(y'_J,\, v'_J)$ belongs to $\Gamma^{\rm
(u)}$. Let us associate a minimizing curve $\gamma_{x,J}$ on the
space-time cylinder $M = \{ 0 \leq y <1, \ -\infty < t < +\infty \}$
with an arbitrary minimizer $\{ y'_j, j \leq J \}$. To construct
$\gamma_{x,J}$ we just connect all the pairs of points $(y'_j, j),
(y'_{j-1}, j-1)$ for $j \leq J$ by straight segments of minimal
length.  Denote by $y'_{x,J}(t)$ a piecewise linear function such that
$\gamma_{x,J} = \{ (y'_{x,J}(t),t), \ -\infty < t < +\infty
\}$. Clearly, $y'_{x,J}(j) = y'_j$.  Denote also by $\gamma$ the
minimizing curve $\{ (x_c,t), \ -\infty < t < +\infty \}$
corresponding to a global minimizer $\{ y_j = x_c \}$.

We have shown above that any minimizing curve $\gamma_{x,J}$ is
asymptotic to $\gamma$ as $t \to -\infty$. By a standard argument this
implies that any two minimizing curves do not intersect each other,
except if they start from the same point $x$. It follows that for all
but at most countably many $x$ and all $J$ there exists a unique
minimizing curve $\gamma_{x,J}$. Denote by $S$ the exceptional set of
$x$'s where a minimizer is not unique. Obviously, for fixed $x$ and
different $J$'s minimizing curves $\gamma_{x,J}$ are connected by a
time shift. It follows that the set $S$ does not depend on $J$.
Hence, for all $x$ outside of $S$ one can define a function
$v(x)\equiv dy'_{x,J}(t)/ dt \ |_{t = J_-} = \pm\rho(y'_J,y'_{J-1})$. It
is easy to see that for any solution $u(x,t),\,\, t \leq T$ to the kicked
Burgers equation in the semi-infinite domain $]-\infty,T]$ one has:
$u(x,k_-) = v(x)$ for all integer $k \leq T$ and $x$ outside of
$S$. This implies that the solution to the kicked Burgers equation in
the semi-infinite domain $]-\infty,t]$ is unique and is generated by
$v(x)$. It also follows that $u_{\infty}(x) = v(x) + g(x)$ is a unique
fixed point for $B_g$. The set $S$ is a set of shocks at integer
moments of time. It follows from the closeness of the set of
minimizers that $v(x)$ is continuous outside of $S$. The
non-intersecting property implies that for arbitrary $x \in S$ there exist
$v(x_-) = \lim_{y \to x_-} v(y),\, v(x_+) = \lim_{y \to x_+} v(y)$,
and that $v(x_-) > v(x_+)$. It also follows that $v(x)$ is a function
of bounded variation. Clearly, $v(x_-)$ and $v(x_+)$ are the
velocities of two minimizing curves which start at $(x,J)$ for any
integer $J$.  Notice that there can be more than two minimizing
curves starting at a shock point $x$; their velocities are between
$v(x_+)$ and $v(x_-)$. This happens, e.g., at shock mergers. Since
$|v(x)| = \rho(y'_J,y'_{J-1})$, we have: $\max_{x}|v(x)| \leq 1/2, \
\max_{x}|u_{\infty}(x)| \leq 1/2 + \max_{x}|g(x)|$. We have seen above
that the points of continuity of $u_{\infty}$, and hence of $v$, are
exactly the points of uniqueness of minimizing curves. Since any
limiting point of a sequence $B_{g}^{n}u_0(x) - g(x)$ gives a velocity
of a minimizing curve starting at $(x,J)$, we have $\lim_{n \to
\infty} (B_{g}^{n}u_0(x) - g(x)) = v(x)$, or $B_{g}^{n}u_0(x) \to
u_{\infty}(x)$ as $n \to \infty$ \ for any $x$ outside of $S$. Suppose
now that there exists another global minimizer $\{ \bar{y}_j \}$. The
same argument as above shows that necessarily $\lim_{j \to
\infty}\,\bar{y}_j = \lim_{j \to -\infty}\,\bar{y}_j = x_c$. If
$\bar{y}_i \neq x_c$ for some $i$, then one can construct a sequence
with smaller action by taking $\tilde{y}_j = x_c, \ |j| \leq j_ 0, \
\tilde{y}_j = \bar{y}_j, \ |j| > j_0$ for sufficiently large
$j_0$. Such construction contradicts to global minimality of $\{
\bar{y}_j \}$ and proves uniqueness of the global
minimizer. Statements 1 - 4 are thus proved.

\subsubsection*{Uniqueness of the main shock}

 Till now we have not used the hyperbolicity properties of the
fixed point $P$. We have seen above that if $v$ is a velocity of a
minimizer which starts at $(x,J)$, then the point $(x,v)$ belongs to
$\Gamma^{\rm (u)}$. Denote by $s$ the natural parameter of length along
$\Gamma^{\rm (u)}$, i.e.
\begin{equation}
\Gamma^{\rm (u)} = (x(s),v(s)), \ s \in \rset, \ (dx/ds)^2 +
(dv/ds)^2 = 1, \ (x(0),v(0))=(x_c,0).
\label{parametrunstable}
\end{equation}
The orientation of $s$ is fixed
by the condition $x(s) \uparrow x(0)$ \ \ as \ $s \uparrow 0$. Define
$(x_j(s),y_j(s)) \equiv {\cal T}^{j}(x(s),v(s)), \ j \leq 0$ and the 
$C^2$ function
\begin{equation}
A(s) \equiv 
\sum _{j \leq 0}\left [{{\rho ^2(y_{j}, y_{j-1})}\over {2}} -
G(y_{j})\right ].
\label{actions}
\end{equation}
The series above converges since $G(x_c) = 0$. It is easy to see that
a point $(x(s),v(s))$ corresponds to a minimizer if and only if $A(s)
= \min_{\tilde{s}} A(\tilde{s})$, the minimum being over all
$\tilde{s}$ such that $x(\tilde{s}) = x(s)$.  Denote by $\bar {A}(x)
\equiv \min_{\tilde{s}} A(\tilde{s})$, the minimum being now taken
over all $\tilde{s}$ such that $x(\tilde{s}) = x$.  Notice that for
any $\epsilon > 0$ there exists $\delta(\epsilon) > 0$ such that $A(s)
> \delta(\epsilon)$ for all $|s| > \epsilon$. Fix $\epsilon$ small
enough, so that $x(s)$ is a monotone function for $|s| \leq
\epsilon$. Now, choose $\epsilon_0$ so small that $\max_{|s| \leq
\epsilon_0}\,{ A(s)} < \delta(\epsilon)$. Then for all $|s| \leq
\epsilon_0$, a point $(x(s),v(s))$ corresponds to the unique minimizer
at $x(s)$.  Hence, we have shown that there are no shocks inside some
neighborhood of $x_c$.

We now construct the main shock. Fix an arbitrary time $t$. Consider
the situation on the space-time cylinder $M$. All minimizing curves
approach the global minimizing curve $\gamma$ either from the right or
from the left (see figure~\ref{f:main}).  Denote by $B_r(t)$ and
$B_l(t)$ the sets of points $x$ on the circle such that there exists a
minimizing curve starting at $(x,t)$ which approaches $\gamma$ from
the right and from the left, respectively. Since minimizing curves do
not intersect, $B_r(t)$ and $B_l(t)$ are closed intervals, \ $B_r(t)
\bigcup B_l(t) = S^1 = [0,1[$ \ and $B_r(t) \bigcap B_l(t)$ consists
of just two points. One of them is $x_c$. Denote the other one $x_{\rm
msh}(t)$. It follows immediately from the construction that $x_{\rm
msh}(t)$ is a shock point; moreover it is the main shock. To prove
uniqueness consider any other shock at time $t = J$ at point
$x$. Then, $(x,v(x_-))$ and $(x,v(x_+))$ either belong both to the
negative-$s$ part of $\Gamma^{\rm (u)}$ or both to its positive-$s$
part; that is,  there exists $s_1,s_2$ such that $(x,v(x_-)) =
(x(s_1),v(s_1))$ and $ (x,v(x_+)) = (x(s_2),v(s_2))$ with $s_1s_2 >
0$. For $j$ sufficiently negative, both $(x_j(s_1),y_j(s_1)) =
{\cal T}^{j}(x(s_1),v(s_1))$ and $ (x_j(s_2),y_j(s_2)) =
{\cal T}^{j}(x(s_2),v(s_2))$ belong to an $\epsilon_0$-neighborhood of
$(x(0),v(0))$ where there is no shock. Hence a prehistory of an
original shock is not longer than $|j|$. Statement 5 is thus proved.

\subsubsection*{Exponential convergence to the unique solution}

Consider a small neighborhood $U$ of the point $P$. It is well known
(Hartman 1960; Belitskii 1973) that if $U$ is small enough then,
inside $U$, ${\cal T}$ is $C^1$-smoothly conjugate to a linear
transformation. This means that there exists a local $C^1$-smooth
change of variables such that in the new coordinates $(X,V)$ the map
${\cal T}$ becomes ${\cal T}_{\lambda} : (X,V) \mapsto (\lambda X,
\lambda^{-1} V)$ and $P$ is the origin in the coordinates $X,V$.
Denote $u_n(x) \equiv B_g^n u_0$ and recall that $\int^{1}_{0}u_0(x)dx
= 0$. For arbitrary $x$ consider a point $(x,v) = (x, u_n(x))$ on the
cylinder ${\cal C}$ and its backward trajectory $(x(-j),v(-j))= {\cal
T}^{-j}(x,v), \ 0 \leq j \leq n$. It is easy to see that there exists
a neighborhood $U_1 \subset U$ such that if two points of this
backward trajectory belong to $U_1$, then all points in between belong
to $U$. Notice, that there exists $n_1$ which depends only on $U_1$
such that, uniformly in $n$, at most $n_1$ points of the backward
trajectory $(x(-j),v(-j)), \ 0 \leq j \leq n,$ are outside of
$U_1$. This implies that for some $j_1, \ 0 \leq j_1 \leq n_1,$ and
$j_2, \ n-n_1 \leq j_2 \leq n,$ we have $(x(-j_1),v(-j_1)) \in U_1,
(x(-j_2),v(-j_2)) \in U_1$. Hence, $(x(-j),v(-j)) \in U$ for all $j_1
\leq j \leq j_2$ and $j_2 - j_1 \geq n - n_1$. Then, there exists a
constant $C_1 > 0$ such that the distance between $(x(-j_1),v(-j_1))$
and a piece of local unstable manifold inside $U$ is less than
$C_1\lambda^{-n}$. Denote by $]s^{(1)},s^{(2)}[, \ s^{(1)} < 0, \
s^{(2)} > 0$ an interval of the parameter which corresponds to this
piece of local unstable manifold. Then the Euclidian distance ${\rm
dist}\,((x(-j_1),v(-j_1)), (x(s_n),v(s_n))) \leq C_1\lambda^{-n}$ for
some $ s_n \in [s^{(1)}, s^{(2)}]$. Denote by ${\bar s}^{(1)}$ and
${\bar s}^{(2)}$ the values of the parameter $s$ corresponding to
${\cal T}^{n_1}(x(s^{(1)}),v(s^{(1)}))$ and ${\cal
T}^{n_1}(x(s^{(2)}),v(s^{(2)}))$, respectively. Then, there exists a
constant $\tilde C > 0$ such that ${\rm dist}\,((x,v), (x(\bar
{s}_n),v(\bar {s}_n))) \leq \tilde C\lambda^{-n}$ for some $\bar{s}_n
\in [{\bar s}^{(1)}, {\bar s}^{(2)}]$.  It might happen that $\bar
{s}_n$ does not correspond to a minimizer.  However, $A(\bar {s}_n)
\to \bar {A}(x)$ as $n \to \infty$. More precisely, one can show that
there exists a constant $C_2 > 0$ such that
\begin{equation} 
A(\bar {s}_n) - \bar {A}(x(\bar{s}_n)) \leq C_2n\lambda^{-n}, \ \ |\bar
{A}(x(\bar{s}_n)) - \bar {A}(x)| \leq C_2\lambda^{-n}.
\label{estimate}
\end{equation}
Suppose now that $u_{\infty}$ is differentiable at $x$ (recall that it
is differentiable almost everywhere). Denote by $s_x$ the value of
parameter $s$ corresponding to the unique minimizer at $x$. Then,
$dx(s)/ds \ |_{s = s_x} \neq 0$. Then, there exist $\epsilon, \
\delta > 0$ such that $|dx(s)/ds| \geq \delta,\,\, |dy(s)/dx|
\leq \delta^{-1}$ for all $s \in ]s_x - \epsilon, s_x + \epsilon[$.  Denote by
$\nu \equiv \min_{s \in [\bar {s}^{(1)}, s_x - \epsilon] \bigcup [s_x +
\epsilon, \bar {s}^{(2)}]} {\rm dist}\, \{ (x, \bar {A}(x)), (x(s), A(s))
\} > 0$. Clearly, there exists $N$ which depends only on $\nu$ such
that ${\rm dist}\, \{ (x, \bar {A}(x)), (x(\bar{s}_n), A(\bar{s}_n)) \}
< \nu$ for all $n > N$. Hence, $\bar{s}_n \in ]s_x - \epsilon, s_x +
\epsilon[$ for all $n > N$. This and the estimates for the derivatives
immediately imply that there exists a constant $C(x) > 0$ such that
$|B_{g}^{n}u_0(x) - u_{\infty}(x)| \leq C(x)\lambda^{-n}$.

To prove an estimate from below, notice that, for any backward trajectory\\
$(x(-j),v(-j))$,  $0 \leq j \leq n$, the last point $(x(-n),v(-n))$
cannot be too close to $P$ if $u_0(x_c) \neq 0$. Indeed, the initial
potential $\psi_0$ has non-zero slope at $x_c$ since we assumed
$u_0(x_c)\neq 0$. Hence, it is possible
to make the action smaller by moving further from $x_c$. It is easy to
show that there exist $\epsilon (u_0) > 0$ such that ${\rm dist}\, \{
(x(-n),v(-n)), (x_c, 0) \} \geq \epsilon (u_0)$. It follows that a
point $(x,v)$ cannot be too close to $(x(s),v(s)), \ s \in [{\bar
s}^{(1)}, {\bar s}^{(2)}]$, i.e.\  there exists a constant $c(u_0) > 0$ such
that $|B_{g}^{n}u_0(x) - u_{\infty}(x)| \geq
c(u_0)\lambda^{-n}$. Statement 6 is thus proved.

\subsubsection*{Analyticity and finiteness of the number of shocks}

To prove Statement 7 we first notice that $\Gamma^{\rm (u)}$ is
analytic, provided $G$ is analytic (see Moser 1956). More precisely,
Moser's result implies local analyticity, from which we can infer
analyticity on any closed interval of $\Gamma^{\rm (u)}$ which does
not contain the fixed point. Denote by $\{ x_i \}$ the (at most
countable) set of shock points other than the main shock at an integer
time $J$.  As we have already seen above, for every $x_i$ there exists
an open set $]s_1(i),s_2(i)[$ of the parameter $s$, where the
parameter values $s_1(i),s_2(i)$ correspond to the points $(x_i,
v({x_{i}}_-))$ and $(x_i, v({x_{i}}_+))$, respectively. Since $x_i$ is
not the main shock, $s_1(i)s_2(i) > 0$. The non-intersecting property
of minimizing curves implies that different intervals
$]s_1(i),s_2(i)[$ do not intersect. Also, $]s_1(i),s_2(i)[ \subset
[s_{2{\rm msh}}, s_{1{\rm msh}}]$, where $s_{1{\rm msh}}$ and
$s_{2{\rm msh}}$ are the parameter values corresponding to $(x_{\rm
msh}, v({x_{\rm msh}}_-))$ and $(x_{\rm msh}, v({x_{\rm msh}}_+))$,
respectively. It is easy to see that all $s \in [s_{2{\rm msh}},
s_{1{\rm msh}}] - \bigcup _{i} (s_1(i),s_2(i))$ correspond to
minimizers. Clearly, for all $i$ there exists $s(i) \in
(s_1(i),s_2(i))$ such that $dx(s)/ds\, |_{s=s(i)} = 0$. Suppose that
there be infinitely many shocks. Then there exists an accumulation
point $s_{\infty}$ for the sequence $\{ s(i) \}$.  It follows that all
derivatives of $x(s)$ vanish at $s_{\infty}$. The analyticity of
$x(s)$ then implies that $x(s)$ is a constant function. This
contradiction proves that the number of shocks is finite. The same
argument works for all times $t$, since for all $t$ the Lagrangian map
transforms any piece of finite length of $\Gamma^{\rm (u)}$ into an
analytic curve. Denote by $\Gamma^{\rm (u)}_t = (x_t(s),\,v_t(s))$ for
$ 0 \leq t \leq 1$ the image of $\Gamma^{\rm (u)}$ under the
Lagrangian map at time $t$, where $s$ is a natural parameter along
$\Gamma^{\rm (u)}_t$. Suppose that the number of non-main shocks at
time $t$ is $K$, so that the total number of shocks is $K+1$. As
above, denote by $]s_1(i),s_2(i)[$ the intervals of the parameter $s$
generated by the $i$-th shock, and by $s_{2{\rm msh}}$ and $s_{1{\rm
msh}}$ the values of the parameter corresponding to the main
shock. The intervals $]s_1(i),s_2(i)[,\, 1 \leq i \leq K$, divide
$[s_{2{\rm msh}}, s_{1{\rm msh}}]$ into $K+1$ closed intervals $I_i,\,
1 \leq i \leq K + 1$. It is easy to see that each of those intervals
corresponds to an analytic piece of $u_{\infty}(x,t)$, i.e.\
$u_{\infty}(x(s),t) = v_t(s),\, s \in I_i$. Finally, we show that the
number of preshock events between time 0 and 1 is finite. Suppose the
number of preshocks were infinite. Denote by $t_i, s_i$ the time of
the $i$-th preshock and the corresponding value of the parameter
$s$. Denote by $(t^*,s^*)$ an arbitrary point of accumulation for a
sequence $(t_i,s_i)$. It is easy to see that all derivatives of
$x_{t^*}(s)$ vanish at point $s^*$. This implies that $x_{t^*}(s)$ is
a constant function.  Again, we get a contradiction, which finishes
the proof of Statement 7.

\end{document}
\newpage
\par\noindent FIGURE CAPTIONS
\vspace{2mm}

\begin{itemize}

\item[Figure 1:] Snapshots of the velocity for the unique time-periodic
solution corresponding to the kicking force $g(x)$ shown in the upper
inset; the various graphs correspond to six output times equally
spaced during one period. The origin of time is taken at a
kick. Notice that during  each period, two new shocks are born  and two
mergers occur.

\item[Figure 2:] Exponential relaxation to a time-periodic solution 
for three different initial velocity data as labelled. The horizontal axis 
gives  the  time elapsed since $t=0$.

\item[Figure 3:] Evolution of shock positions during one period. The
beginnings of lines correspond to births of shocks (preshocks) at
times $t_{\star 1}$ and $t_{\star 2}$; shock mergers take place at
times $t_{c1}$ and $t_{c2}$. The ``main shock'', which survives for
all time, is shown with a thicker line.

\item[Figure 4:] Sketch of a hyperbolic fixed point $P$ with stable
($\Gamma ^{\rm (s)}$) and unstable ($\Gamma ^{\rm (u)}$)
manifolds. The dashed line gives the orbit of successive iterates of a
point near the stable manifold.

\item[Figure 5:] Unstable manifold $\Gamma ^{\rm (u)}$ on the $(x,v)$-cylinder
(the $x$-coordinate is defined modulo~1) which passes through the
fixed point $P=(x_c,0)$. The bold line is the graph
of $u_\infty(x,1_-)$. The main shock is located at
$x_l=x_r$. Another shock at $x_1$ corresponds to a local zig-zag of
$\Gamma ^{\rm (u)}$ between A and B.

\item[Figure 6:] Minimizers (trajectories of fluid particles) on the
$(x,t)$-cylinder. Time starts at $-\infty$. Shock locations at $t=0_-$
are characterized by  having two minimizers (an instance is at $x_1$).
The main shock is at $x_l=x_r$. The fat line $x=x_c$ is the global minimizer.

\item[Figure 7:] Pdf of the velocity gradient at negative values in log-log
coordinates. Upper inset: local scaling exponent. A power law with
exponent $-7/2$ is obtained at large arguments.

\item[Figure 8:] Same as figure~\ref{f:first-deriv} with the second space
derivative of the velocity. The exponent is now $-2$.

\item[Figure 9:] Upper part: Pdf of (negative) velocity increments in
log-log coordinates for various values of the separation $\Delta x$
in geometric progression from $2\pi/N$ to $2^7 (2\pi/N)$. Lower part:
the corresponding local scaling exponents.

\item[Figure 10:] Structure functions $S_p(\Delta x)$ for $p=2,3,4,5$ as
labelled. Note the linear behavior at small $\Delta x$.

\item[Figure 11:] Merger of two shocks $X_{kl}(t)$ and $X_{kr}(t)$ on the
$(x,t)$-cylinder. The merger (event $M_k$)  takes place at $t=\tilde t_k$.
At the earlier time $t=\theta_k$ the shocks are within a distance
$\Delta x$.

\end{itemize}


\begin{thebibliography}{99}

\bibitem{aubry}
Aubry, S.
1983.\,\,
The twist map, the extended Frenkel--Kontorova model and the devil's
staircase,
{\sl Physica {\rm D} \bf 7}, 240--258.

\bibitem{BarabasiStanley}
{Barab\'asi,~A.-L. \& Stanley,~H.E.}
{1995}.\,\,
{\it Fractal Concepts in Surface Growth},
{Cambridge University Press},
{Cambridge}.

\bibitem{BecFrisch}
{Bec,~J. \& Frisch,~U.}
{2000}.\,\,
{Pdf's of derivatives and increments for decaying Burgers
turbulence}, {\sl Phys. Rev.} E{\bf 61}, 1395--1402. (cond-mat/9906047).

\bibitem{BecFrischVillone}
{Bec,~J., Frisch,~U.} \& Villone,~B. 
{2000}.\,\,Singularities and  the distribution of density in the
Burgers/adhesion model, submitted to {\sl Physica }D. (cond-mat/9912110)

\bibitem{belitskii}
Belitskii,~G.R. 1973.\,\,
Functional equations and conjugacy of local
diffeomorphisms of a finite smoothness class,
{\sl Funct.  Analys.  Applic. \bf 7}, 268--277.
(Translated from Russian: Funktsional'nyi Analiz i Ego Prilozheniya
{\bf 7} (1973), 17--28).

\bibitem{Boldyrev}
Boldyrev,~S.-A. 1997.\,\, Velocity-difference probability
density functions for Burgers turbulence, {\sl Phys.~Rev.~{\rm E}
\bf 55}, 6907--6910.

\bibitem{BouchaudMezardParisi}
Bouchaud,~J.-P., M\'ezard,~M. \& Parisi,~G. 1995.\,\, Scaling and
intermittency in Burgers turbulence, {\sl Phys.~Rev.~{\rm E}
\bf 52}, 3656--3674.

\bibitem{ChekhlovYakhot95}
Chekhlov,~A. \& Yakhot,~V. 1995.\,\, Kolmogorov turbulence in a
random-force-driven Burgers equation: anomalous scaling and
probability density functions, {\sl Phys.~Rev.~{\rm E}
\bf 52}, 5681--5684.

\bibitem{Cole}
Cole,~J.D. 1951.\,\, On a quasi-linear parabolic equation
occurring in aerodynamics, {\sl Quart. Appl. Math. \bf 9},
225--236.

\bibitem{crighton-scott}
Crighton,~D.G. \& Scott,~J.F. 1979.\,\,
Asymptotic solutions of model equations in nonlinear acoustics,
{\sl  Phil. Trans. Roy. Soc. London \bf 292}, 101--134.

\bibitem{weinane}
E,~W. 1999.\,\,
Aubry--Mather theory and periodic solutions for the forced
Burgers equation,
{\sl Comm. Pure Appl. Math. \bf 52}, 0811--0828.

\bibitem{EKMS97}
E,~W., Khanin,~K., Mazel,~A. \& Sinai,~Ya. 1997.\,\,
Probability distribution functions for the random forced
Burgers equation, {\sl Phys. Rev. Lett. \bf 78}, 1904--1907.

\bibitem{EKMS98}
E,~W., Khanin,~K., Mazel,~A. \& Sinai,~Ya. 2000.\,\,
Invariant measures for Burgers equation with stochastic forcing,
{\sl Ann. Math.} in press.

\bibitem{EVandenE1} E,~W. \& Vanden~Eijnden,~E. 1999.\,\, Asymptotic
theory for the probability density functions in Burgers turbulence,
{\sl Phys. Rev. Lett. \bf 83}, 2572-2575. (chao-dyn/9901006)

\bibitem{EVandenE2} E,~W. \& Vanden~Eijnden,~E. 2000.\,\,
Statistical theory for the stochastic Burgers equation in the inviscid
limit, {\sl Comm. Pure Appl. Math.}, in press.

\bibitem{cup}
Frisch,~U. 1995.\,\,
{\it Turbulence\,: the Legacy of A.N.~Kolmogorov},
Cambridge University Press,
Cambridge.

\bibitem{Gotoh}
Gotoh,~T. 1999.\,\,
Probability density functions in steady-state Burgers turbulence,
{\sl Phys.~Fluids.\bf 11}, 2143--2148.


\bibitem{GotohKraichnan}
{Gotoh,~T. \& Kraichnan,~R.H.} {1998}.\,\,
{Steady-state Burgers turbulence with large-scale forcing},
{\sl Phys.~Fluids.\bf 10}, {2859--2866}.

\bibitem{hartman}
Hartman,~P. 1960.\,\,
On local homeomorphisms of Euclidean spaces,
{\sl Bol. Soc. Mat. Mexicana} (2) {\bf 5}, 220--241.

\bibitem{Henon}
H\'enon,~M. 1983,
Numerical exploration of Hamiltonian systems, in 
{\it Chaotic behaviour of deterministic systems, Les Houches 1981}
G.~Iooss, R.~Helleman \& R.~Stora, eds., pp.~53--170, North Holland.

\bibitem{Hopf}
Hopf,~E. 1950.\,\, The partial differential equation
$u_t+uu_x=u_{xx}$, {\sl Comm. Pure Appl. Mech. \bf 3}, 201--230.

\bibitem{jauslin-kreiss-moser}
Jauslin,~H.R., Kreiss,~H.O. \& Moser,~J.
1997.\,\,
On the forced Burgers equation with periodic boundary conditions, 
preprint.

\bibitem{KardarParisiZhang}
{Kardar,~M., Parisi,~G. \&\,\,Zhang,~Y.-C.}
{1986}.\,\,
{Dynamical scaling of growing interfaces},
{\sl Phys.~Rev.~Lett. \bf 56}, {889--892}.

\bibitem{kolmogorov}
{Kolmogorov,~A.N.}
{1941}.\,\,
{Dissipation of energy in locally isotropic turbulence},
{\sl {Dokl. Akad. Nauk SSSR}}
{\bf {32}}, {16--18}
{({reprinted in {\sl Proc. R. Soc. Lond.} A {\bf 434}, 15--17 (1991)})}.

\bibitem{Kraichnan}
Kraichnan,~R.H. 1999.\,\, Note on forced Burgers turbulence,
{\sl Phys.~Fluids \bf 11}, 3738-3742. (chao-dyn/9901023)

\bibitem{lax}
Lax,~P.-D. 1957.\,\, Hyperbolic systems of conservation laws II,
    {\sl Comm. Pure Appl. Math. \bf 10}, 537--566.

\bibitem{Manneville}
Manneville,~P. 1990,
{\it Dissipative Structures and Weak Turbulence},
Perspectives in Physics, Academic Press. 

\bibitem{mather}
Mather,~J.N. 1982.\,\,
Existence of quasi-periodic orbits for twist homeomorphisma of the
annulus,
{\sl Topology \bf 21}, 457--467.

\bibitem{moser}
Moser,~J. 1956.\,\,
The analytic invariants of an area-preserving mapping near
a hyperbolic fixed point,
{\sl Comm. Pure Appl. Math. \bf 9}, 673--692.

\bibitem{newman}
Newman,~T.J. \& McKane,~A.J. 1997.\,\, Directed lines in sparse
potentials, {\sl Phys. Rev.  {\rm E} \bf 55}, 165--175.


\bibitem{NoullezVergassola}
Noullez,~A. \& Vergassola,~M. 1994.\,\, A fast algorithm for
discrete Legendre transforms, {\sl J. Sci. Comp. \bf 9},
259--281.

\bibitem{oleinik}
Oleinik,~O. 1957.\,\,
Discontinuous solutions of nonlinear differential
equations,
{\sl Uspekhi Mat. Nauk \bf 12}, no~3, 3--73. 
({\sl Russ. Math. Survey.}, Amer. Math. 
Transl. Series 2 {\bf 26}, 95--172).

\bibitem{Polyakov}
Polyakov,~A.M. 1995.\,\, Turbulence without pressure,
    {\sl Phys. Rev.  {\rm E} \bf 52}, 6183--6188.

\bibitem{SheAurellFrisch}
{She,~Z.S., Aurell,~E. \& Frisch,~U.}
{1992}.\,\,
{The inviscid Burgers equation
with initial data of Brownian type},
{\sl Commun.~Math.~Phys.}
{\bf {148}}, {623--641}.

\bibitem{sobolevski}
Sobolevski,~A.~N. 1999\,\,
On periodic solutions of
a Hamilton--Jacobi equation with periodic forcing, 
{\sl Mat.\ Sbornik} (to appear, in Russian). Also chao-dyn/9906035 in English.

\bibitem{trussov}
Trussov,~A. 1996.\,\,
Linear time algorithms for discrete Legendre transform and for the
numerical simulation of Burgers' equation, {\sl Computational
Seismology \bf  29},  45--53.

\bibitem{VergassolaDubrulleFrischNoullez}
{Vergassola,~M., Dubrulle,~B., Frisch,~U. \& Noullez,~A.}
{1994}.\,\,
{Burgers' equation, Devil's staircases and the mass distribution
for large-scale structures},
{\sl Astron.~Astrophys.}
{\bf {289}}, {325--356}.

\end{thebibliography}
\end{document}